\documentclass[%
 reprint,
 superscriptaddress,
 amsmath,amssymb,
 aps,
]{revtex4-2}

\usepackage{graphicx}
\usepackage{dcolumn}
\usepackage{bm}
\usepackage{upgreek}

\usepackage{color} 
\usepackage{isomath} 
\usepackage[version=4]{mhchem} 

\newcommand{\D}{\text{D}}
\newcommand{\B}{\text{B}}

\newcommand{\p}{\text{p}}
\renewcommand{\r}{\text{r}}
\newcommand{\f}{\text{f}}
\newcommand{\RE}{\text{Re}}
\newcommand{\St}{\text{St}}
\newcommand{\mw}{\text{mw}}
\newcommand{\cm}{\text{cm}}

\newcommand{\e}{\text{e}}
\newcommand{\cc}{\text{c}}
\newcommand{\h}{\text{h}}

\begin{document}

\preprint{APS/123-QED}

\title{Quincke oscillations of colloids at planar electrodes}

\author{Zhengyan Zhang}
\affiliation{Department of Chemical Engineering, Columbia University, New York, NY 10025, USA}

\author{Hang Yuan}
\affiliation{Applied Physics Program, Northwestern University, Evanston, IL 60208, USA}

\author{Yong Dou}
\affiliation{Department of Chemical Engineering, Columbia University, New York, NY 10025, USA}

\author{Monica Olvera de la Cruz}
\email{m-olvera@northwestern.edu}
\affiliation{Department of Materials Science and Engineering, Northwestern University, Evanston, IL 60208, USA}
\affiliation{Applied Physics Program, Northwestern University, Evanston, IL 60208, USA}

\author{Kyle J. M. Bishop}
\email{kyle.bishop@columbia.edu}
\affiliation{Department of Chemical Engineering, Columbia University, New York, NY 10025, USA}

\date{\today}

\begin{abstract}
Dielectric particles in weakly conducting fluids rotate spontaneously when subject to strong electric fields. Such Quincke rotation near a plane electrode leads to particle translation that enables physical models of active matter. Here, we show that Quincke rollers can also exhibit oscillatory dynamics, whereby particles move back and forth about a fixed location. We explain how oscillations arise for micron-scale particles commensurate with the thickness of a field-induced boundary layer in the nonpolar electrolyte. This work enables the design of colloidal oscillators.
\end{abstract}

\maketitle



Solid particles in weakly conducting fluids are long known to rotate spontaneously when subject to strong electric fields  \cite{Quincke1896}. So-called Quincke rotation near a solid boundary enables particle propulsion underlying recent experimental models of active matter \cite{Bricard2013,Bricard2015, Zhang2020}. The mechanism of Quincke rotation is well described by the Taylor-Melcher leaky dielectric model, which treats the fluid as a homogeneous Ohmic conductor containing no free charge \cite{Melcher1969, Saville1997}. For nonpolar electrolytes \cite{Prieve2017} subject to strong fields, the validity of this assumption requires the rapid generation and recombination of charge carriers within the fluid. To maintain an  electric current, carriers must be generated within fluid volumes of finite thickness near system boundaries.  Within such boundary layers, the assumption of the leaky dielectric model breaks down, and new types of electrohydrodynamic phenomena can arise.

For a symmetric binary electrolyte, the boundary layer thickness can be approximated as $\ell = e \mu E_{\e}/k_{\r} n_o$ where $e$, $\mu$, and $n_o$ are the charge, mobility, and density of carriers, $E_{\e}$ is the external field strength, and $k_{\r}$ is a rate constant for ion recombination \cite{supp}. Carriers are removed from the boundary region at a rate equal to the flux $e \mu n_o  E_{\e}$. At steady-state, this flux is balanced by carrier generation within the boundary layer, which occurs at a rate equal to that of carrier recombination in the bulk $k_{\r} n_o^2$. For nonpolar solutions of AOT surfactant commonly used in the study of Quincke rollers, external fields are expected to generate boundary layers as large as 10 $\mu$m---comparable to the size of colloidal particles.

Here, we investigate the dynamics of particles within such field-induced boundary layers and observe oscillatory motions that are not predicted by the leaky dielectric model. Our experiments are based on polystyrene spheres dispersed in AOT-hexadecane solutions above a planar electrode. The application of an electric field above a critical magnitude causes the particles to roll steadily across the electrode surface \cite{Bricard2013}. Upon further increasing the field strength, however, the particles begin to oscillate back-and-forth with an amplitude comparable to their diameter. Owing to their small size, the oscillations cannot be attributed to inertial effects.  Moreover,  simulations based on the leaky dielectric model are unable to reproduce the observed oscillations---even when accounting for the proximal electrode.  By relaxing model assumptions to account for the finite rates of ion formation and recombination, we show how Quincke oscillations can arise for particles comparable in size to the boundary layer thickness. Oscillations derive from a memory effect caused by the anisotropic charging of the particle surface. Consistent with this mechanism, we demonstrate that oscillations are not observed for larger particles that extend beyond the boundary layer or for particles moving within the bulk electrolyte. Together, these results enable the design of colloidal oscillators and highlight the significance of electric boundary layers on the active motions of particles and their ensembles. 

\begin{figure}[t]
    \centering
    \includegraphics[width=8cm]{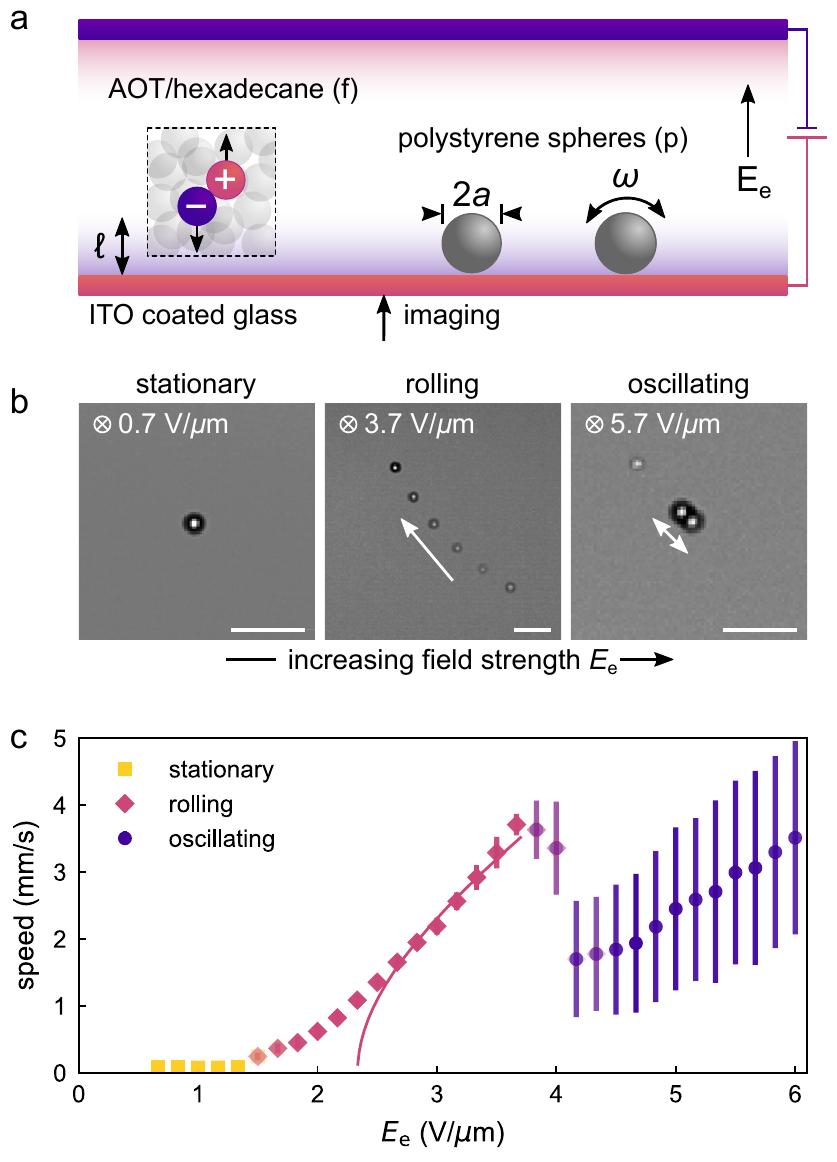}
    \caption{(a) Schematic illustration of the experimental setup. 
    (b) Time-lapse microscopy images showing the three observed particle behaviors: stationary, rolling, and oscillating. Here, the particle radius is $a=5~\mu$m, the AOT concentration is $[\text{AOT}]=150$ mM, and the electrode separation is $L=150~\mu$m. Scale bars are 40 $\mu$m. (c) Time-averaged particle speed vs.\ external field strength $E_{\e}$. For each 20 ms trajectory, we compute the mean and standard deviation of the particle speed. Markers denote the median of these mean speeds for ca.~1000 trajectories; error bars denote the median of the corresponding standard deviations. The plotted data are colored based on probability assignments of the Bayesian classifier. The solid curve is a fit of the form $U = (\kappa a/\tau_{\mw})[(E_{\e}/E_{o})^2-1]^{1/2}$ with $\kappa=0.40$ and $E_{o} = 2.3$ V/$\mu$m; the Maxwell-Wagner time is $\tau_{\mw}=0.70$ ms from independent conductivity measurements \cite{supp}. Note that the fitted value of the field strength $E_o$  differs from that predicted by equation (\ref{eq:Ec}) for an unbounded sphere, $E_{\cc}=0.91$ V/$\mu$m.}
    \label{fig:f1}
\end{figure}

In our experiments, polystyrene spheres are dispersed at low volume fraction in hexadecane solutions of AOT surfactant. The dispersion is sandwiched between parallel electrodes, where the particles sediment under gravity to the lower boundary (Fig.\ \ref{fig:f1}a).  Upon application of an external field $E_{\e}$, the particles move on the electrode surface as captured by high speed video microscopy. Depending on the strength of the applied field, we observe three types of particle motion termed stationary, rolling, and oscillating (Fig.\ \ref{fig:f1}b,c). 

For external fields weaker than a critical value, particles remain motionless (Fig.\ \ref{fig:f1}b, left). Above this value, particles roll along the electrode in random directions perpendicular to the applied field with a constant speed (Fig\ \ref{fig:f1}b, middle). Further increasing the field, we observe a second transition whereby particles cease to roll and instead oscillate back-and-forth (Fig.\ \ref{fig:f1}b, right). The time-averaged particle speed increases with field strength before slowing abruptly at the onset of oscillations (Fig.\ \ref{fig:f1}c, markers). Accompanying this transition from rolling to oscillating, temporal variations in particle speed increase in magnitude from zero to a finite value (Fig.\ \ref{fig:f1}c, error bars). In addition to these descriptive statistics, we use Bayesian model selection \cite{sivia2006data,supp} to classify each particle trajectory based on competing models for stationary, rolling, and oscillating dynamics (Fig.\ \ref{fig:f1}c, colors).

The observed transition from stationary to rolling agrees qualitatively with predictions of the leaky dielectric model for a spherical particle immersed in an unbounded fluid with respective permittivities $\varepsilon_{\p}$, $\varepsilon_{\f}$ and conductivities $\sigma_{\p}$, $\sigma_{\f}$. The model predicts that the stationary solution becomes unstable when the external field strength exceeds the critical value \cite{Jones1984, Das2013, Hu2018}
\begin{equation}
    E_{\cc} = \sqrt{\frac{2\eta}{\varepsilon_{\f}\tau_{\mw}(\varepsilon_{\cm}-\sigma_{\cm})}} \label{eq:Ec}
\end{equation}
where $\eta$ is the fluid viscosity, $\tau_{\mw}=(\varepsilon_{\p}+2\varepsilon_{\f})/(\sigma_{\p}+2\sigma_{\f})$ is the Maxwell-Wagner time, and $x_{\cm}=(x_{\p}-x_{\f})/(x_{\p}+2x_{\f})$ for $x=\varepsilon,\sigma$ are the Claussius-Mossotti factors characterizing the high and low-frequency polarizability of the sphere, respectively. Above this field, the angular velocity and thereby the rolling speed $U$ increase with increasing field strength as $U = (\kappa a/\tau_{\mw})[(E_{\e}/E_{\cc})^2-1]^{1/2}$ where $\kappa\leq 1$ is a dimensionless coefficient characterizing the strength of rotation-translation coupling (Fig.\ \ref{fig:f1}c, solid curve).  Consistent with this model, the critical field strength is independent of particle radius $a$ but increases with increasing AOT concentration, which increases the conductivity of the fluid \cite{supp}.

Near the transition from rolling to oscillating, particles exhibit a mixture of intermediate behaviors such as rolling in a common direction with a time-periodic speed and rolling with aperiodic reversals in direction \cite{supp}. Similar behaviors attributed to inertial effects were reported for larger spheres ($a = 50~\mu$m) under stronger confinement ($L/a \approx 4$) \cite{pradillo2019quincke}. Here, we neglect this transition region and focus instead on the previously unreported phenomenon of back-and-forth oscillations.

\begin{figure}[t]
    \centering
    \includegraphics{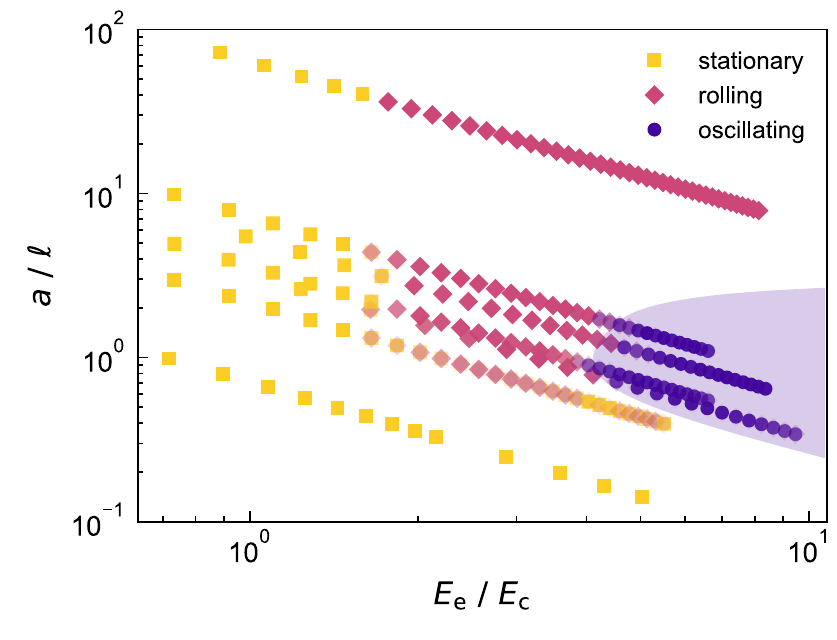}
    \caption{Phase diagram showing the observed dynamics as a function of two dimensionless parameters: $a/\ell$, the ratio of the particle radius and the boundary layer thickness; $E_{\e} / E_{\cc}$, the ratio of the external field strength and the critical field of equation (\ref{eq:Ec}). Plotted data correspond to experiments on five different particle sizes $a=0.5, 1.5, 2.5, 5, 25~\mu$m (for $[\text{AOT}]=150$ mM) and three different AOT concentrations $[\text{AOT}]=50,100,150$ mM (for $a=5~\mu$m). Markers are colored based on probability assignments of the Bayesian classifier \cite{supp}.}
    \label{fig:phase_diagram}
\end{figure}

Oscillatory dynamics are reliably observed for strong fields, $E_{\e}/E_{\cc}>3$, when the ratio between the particle radius and the boundary layer thickness is of order unity, $a/\ell \sim 1$ (Fig.\ \ref{fig:phase_diagram}). In estimating this length scale, $\ell = e\mu E_{\e}/k_{\r} n_o$, we approximate the mobility of AOT micelles as $\mu=(6\pi \eta a_{\h})^{-1}$ where $a_{\h}=1.7$ nm is the reported hydrodynamic radius \cite{Kotlarchyk1985}. We further assume that the rate constant for neutralizing collisions among charged micelles is diffusion-limited such that $k_{\r} = 2 e^2 \mu/\varepsilon_{\f}$ \cite{Debye1942,Saville1997}. Finally, we estimate the concentration of charged micelles from the measured conductivity as $n_o = \sigma_{\f} /2 e^2 \mu$ \cite{supp}. The resulting boundary layer thickness $\ell$ varies from 1 to 20 $\mu$m depending on the AOT concentration and the external field strength. Notably, large particles ($a/\ell \gg 1$) that extend beyond the boundary region do not oscillate but rather roll at even the highest fields investigated (Fig.\ \ref{fig:phase_diagram}). Small particles ($a/\ell \ll 1$) do not move at all; their otherwise Brownian motion is arrested upon application of the field \cite{supp}.

\begin{figure}[t]
    \centering
    \includegraphics[width=8.5cm]{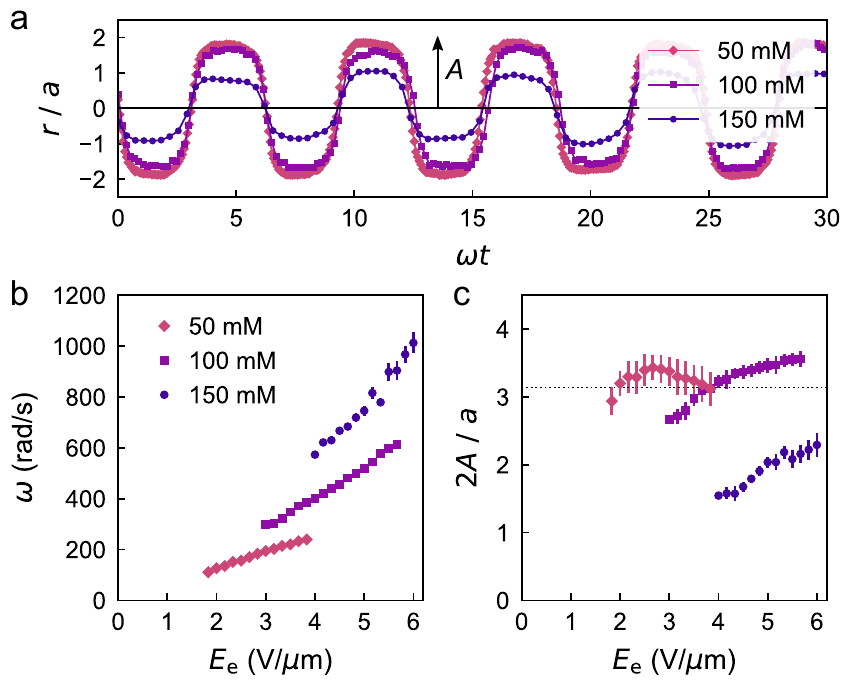}
    \caption{(a) Particle position $r$ vs.\ time $t$ for PS spheres (radius $a=5~\mu$m) in different AOT-hexadecane solutions. The applied field is $E_{\e}/E_{\cc} = 5.4$, which corresponds to $E_{\e}=2.2$, 3.7, and 4.8 V/$\mu$m for $[\text{AOT}]=50$, 100, and 150 mM, respectively. (b) Oscillation frequency $\omega$ vs.\ external field strength $E_{\e}$ for different AOT concentrations. Markers denote the mean frequencies within populations of particle trajectories of equal duration; error bars denote standard deviations of these populations.
    (c) Peak-to-peak oscillation amplitude $2A$ vs.\ external field strength $E_{\e}$ for the three AOT concentrations in (b). Markers denote the mean frequencies; error bars denote standard deviations.}
    \label{fig:oscillating}
\end{figure}

The frequency of particle oscillations $\omega$ is comparable to the dipolar relaxation rate $\tau_{\mw}^{-1}$ and increases with increasing field strength (Fig.\ \ref{fig:oscillating}a,b). Experiments at different AOT concentrations suggest that the oscillation frequency is well approximated as $\omega \approx 0.09 \tau_{\mw}^{-1} E_{\e}/E_{\cc}$ \cite{supp}. This form is identical to that of the rolling frequency predicted by the leaky dielectric model, suggesting that the oscillation frequency is set by a similar balance of particle rotation and charge accumulation at the particle surface.  

The peak-to-peak amplitude of the oscillating particle position is approximately $2A \approx \pi a$ (Fig.~\ref{fig:oscillating}c). This observed quantity is linearly related to the angle $2 A/ \kappa a$ by which the particle rotates during each half of the oscillation cycle. If one assumes frictional rolling with $\kappa=1$, the observed amplitude would imply a rotation of ca.~180$^{\circ}$. By contrast, the assumption of hydrodynamic rolling with a thin lubricating film \cite{Kim2005} requires that $\kappa\leq 1/4$ and implies a rotation of at least two revolutions per half cycle. Below, we present a model for particle oscillations that favors the former interpretation based on frictional rolling.

Owing to the small size of the particles, the observed oscillations cannot be attributed to inertial effects. The Reynolds number for particle oscillations is much less than unity, $\RE = \rho \omega a^2/\eta \sim 10^{-3}$, where $\rho$ is the fluid density. The hydrodynamic resistance to motion is therefore proportional to the particle velocity.  Moreover, particle inertia is also negligible as evidenced by the small Stokes number, $\St = \rho_{\p} a^2 / 15 \eta \tau_{\mw}\sim 10^{-4}$, where $\rho_{\p}$ is the density of the particle. With finite particle inertia, Quincke dynamics of a sphere in an unbounded fluid is mathematically identical to the Lorenz system \cite{Lorenz1963} and to the Malkus water wheel \cite{Kolar1992}, which are known to exhibit oscillatory and chaotic dynamics \cite{Peters2005}. In the absence of inertial effects, however, only the stationary and rolling solutions are permitted by the leaky dielectric model in an unbounded fluid.

Control experiments on particles within the bulk fluid suggest that oscillatory dynamics occur only near the electrode surface. We use a standing acoustic field to levitate particles at the mid-plane between two planar electrodes \cite{sabrina2018shape} and observe their motion upon application of the electric field \cite{supp}.  In the absence of the acoustic field, the application of a strong electric field drives the particles to oscillate at the electrode surface. Such oscillations are not observed when the same field is applied to particles levitating at the mid-plane of the chamber. Instead, particles in the bulk fluid exhibit steady rotation consistent with predictions of the leaky dielectric model.

\begin{figure}[t!]
    \centering
    \includegraphics{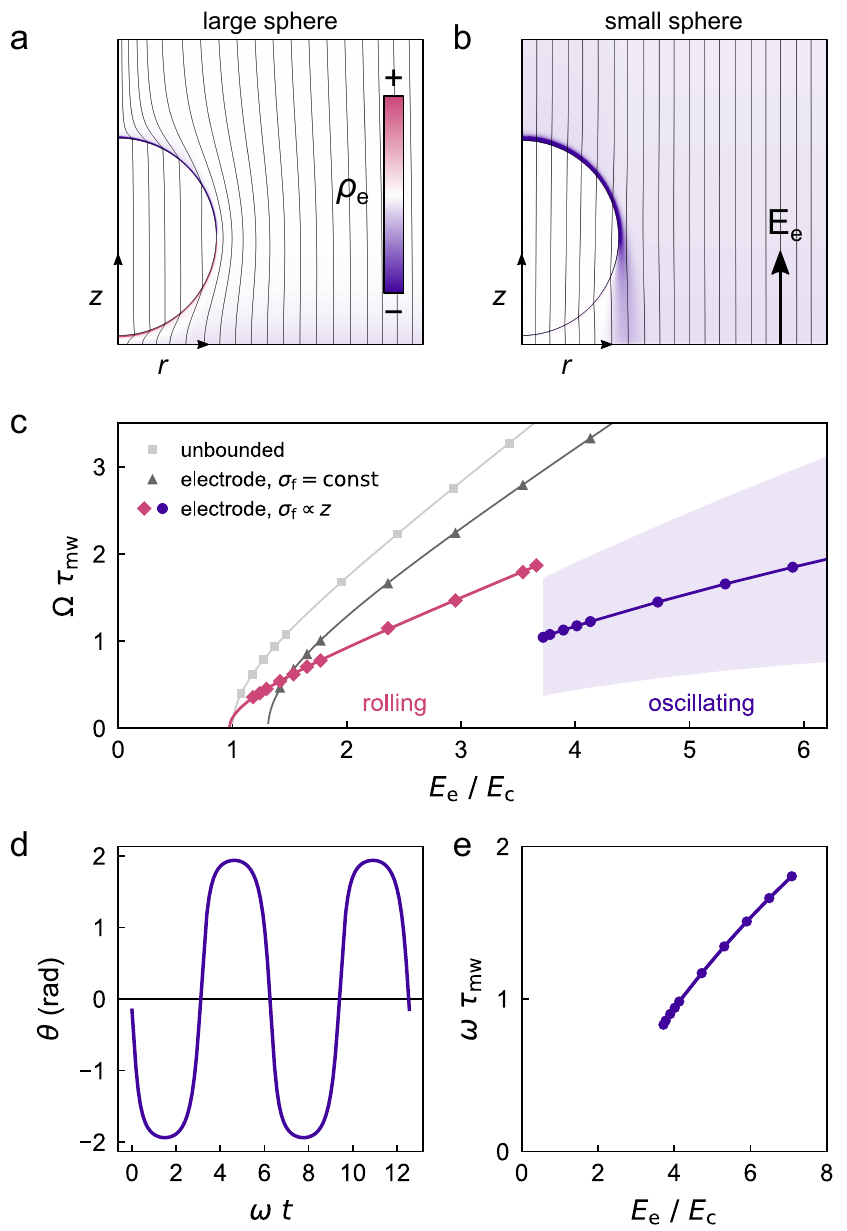}
    \caption{(a,b) Simulated electric field around a stationary sphere in a model electrolyte above a plane electrode; color map shows the charge density \cite{supp}. The radii of the large (a) and small (b) spheres are $a/\ell=3.5$ and $a/\ell=0.14$, respectively. Other parameters include the Debye length $\lambda_{\D}/\ell = 0.028$, the surface separation $\delta/a=0.1$, the particle permittivity $\varepsilon_{\p}/\varepsilon_{\f}=1.2$, and the recombination rate constant $k_{\r}\varepsilon_{\f}/e^2\mu=2$. (c) Time-averaged angular speed $\Omega$ scaled by  $\tau_{\mw}^{-1}$ vs.\ external field strength $E_{\e}$ scaled by $E_{\cc}$ for three variations of the leaky dielectric model: an unbounded sphere, a sphere at a plane electrode with constant fluid conductivity, and a sphere at an electrode with a conductivity gradient. The particle permittivity is $\varepsilon_{\p}/\varepsilon_{\f}=1.5$; the surface separation is $\delta/a=0.1$; the conductivity gradient is $\sigma_{\f} /(2a+\delta)$; the resistance coefficient is $R/8\pi\eta a^3 = 1.45$. The shaded region denotes one standard deviation about the average speed. (d) Angular position $\theta$ vs.\ oscillation phase $\omega t$ for $E_{\e}=5.3 E_{\cc}$. (e) Oscillation frequency $\omega$ scaled by $\tau_{\mw}^{-1}$ vs.\ external field strength $E_{\e}$ scaled by $E_{\cc}$.}
    \label{fig:model}
\end{figure}

To understand why particles of intermediate size oscillate near the electrode (see Fig.\ \ref{fig:phase_diagram}), we first consider the transport of charged AOT micelles around a stationary sphere near a plane boundary (Fig.\ \ref{fig:model}a,b). The electric field and the carrier densities are modeled using the Poisson-Nernst-Planck (PNP) equations modified to describe the generation and recombination of charged micelles within the electrolyte \cite{Saville1997, peters2009size, supp}.  At steady-state, the solution is characterized by three length scales: the particle radius $a$, the Debye length $\lambda_{\D}=(\varepsilon_{\f} k_{\B} T/2e^2 n_o)^{1/2}$, and the boundary layer thickness $\ell$ associated with carrier recombination. We focus our analysis on the limit of strong fields relevant to our experiments, for which $E_{\e}\gg k_{\B} T/ e \ell$ or, equivalently, $\ell\gg \lambda_{\D}$. Under these conditions, the behavior of large spheres ($a\gg\ell$) is well described by the leaky dielectric model.  Charge accumulates at the particle surface as to redirect the electric field and the associated electric current around the particle (Fig.\ \ref{fig:model}a). The Quincke instability is caused by the relaxation of this dipolar charge distribution via particle rotation.

For small spheres ($a\ll\ell$), however, the accumulation of charge at the particle surface is mitigated by the diffusive-leaking of charge carriers around the sides (Fig.\ \ref{fig:model}b). The comparatively little charge that accumulates does not significantly alter the electric field. Without a sufficiently large dipole moment directed antiparallel to the external field, there can be no Quincke rotation for these small particles (cf.\ Fig.\ \ref{fig:phase_diagram}). Moreover, such particles are characterized by a net charge that contributes additional electrostatic forces directed to the nearby electrode. The attraction of small particles to the electrode surface helps to explain the field-induced arrest of their Brownian motion.

For particles of intermediate size ($a\sim\ell$), the observed oscillations are explained by asymmetries in the rates of charging between the top and bottom of the particle. Within the confined region separating the particle and the electrode, ionic currents are limited by the finite rate of ion formation in the fluid. The effective conductivity within such a region of thickness $\delta$ can be approximated as $e k_{\r}n_o^2 \delta / E_{\e}$, which is smaller than the bulk conductivity, $\sigma_{\f}\approx 2 e^2 \mu n_o$, by a factor of $\ell/\delta\gg 1$ \cite{supp}. By modifying the leaky dielectric model to describe variations in the effective conductivity as a function of distance from the electrode surface, numerical simulations are able to reproduce the particle oscillations observed in experiment (Fig.\ \ref{fig:model}c, pink diamonds \& blue circles). 

In the model, we consider a dielectric sphere of radius $a$ immersed in a conductive fluid at a distance $\delta$ from a plane electrode. Application of an external field $E_{\e}$ drives the accumulation of charge at the particle-fluid interface; the effects of free charge within the electrolyte are neglected. The fluid conductivity is assumed to vary with distasnce $z$ from the electrode as $\sigma_{\f} z/(2 a+\delta)$, approaching the bulk value $\sigma_{\f}$ at the top of the particle. The angular velocity of the particle (parallel to the plane) is linearly related to the electric torque as $\Omega=L/R$, where $R=8\pi\eta a^3 f(\delta/a)$ is the relevant resistance coefficient. With these assumptions, the particle dynamics agree qualitatively with the experimental observations (cf.\ Figs.\ \ref{fig:f1}c \& \ref{fig:model}c). At sufficiently high field strengths---here, greater than 3.7 times the critical field $E_{\cc}$ for an unbounded sphere---the particle oscillates back and forth with an peak-to-peak amplitude of ca.\ 200$^{\circ}$ (Fig.\ \ref{fig:model}d). The oscillation frequency $\omega$ increases in proportion to the external field strength $E_{\e}$ (Fig.\ \ref{fig:model}e).

Physically, particle oscillations combine the basic elements of the traditional Quincke mechanism---namely, charge accumulation and mechanical relaxation---with an added memory effect caused by anisotropic charging within the electric boundary layer. Additional experiments on particles of different shapes suggest that these Quincke oscillations can be achieved for any dielectric particle of suitable size \cite{supp}. This mechanism may therefore provide a useful experimental model for active matter \cite{marchetti2013hydrodynamics} comprised of many self-oscillating units, where particle interactions---neglected herein---mediate their collective dynamics. More generally, Quincke oscillations illustrate the potential importance of field-induced boundary layers within nonpolar fluids. Even away from electrode surfaces, such boundary layers are expected to influence the dynamics of micron-scale Quincke swimmers moving within bulk fluids \cite{das2019active, zhu2019propulsion, sherman2020spontaneous}.

\begin{acknowledgments}
This work was supported as part of the Center for Bio-Inspired Energy Science, an Energy Frontier Research Center funded by the U.S. Department of Energy, Office of Science, Basic Energy Sciences under Award DE-SC0000989. 
\end{acknowledgments}

\bibliography{main_refs}

\end{document}


\maketitle

\clearpage
\tableofcontents

\clearpage
\section{AOT-Hexadecane Electrolyte}

\subsection{Conductivity Measurements}

\begin{figure}[h]
    \centering
    \includegraphics[width=13cm]{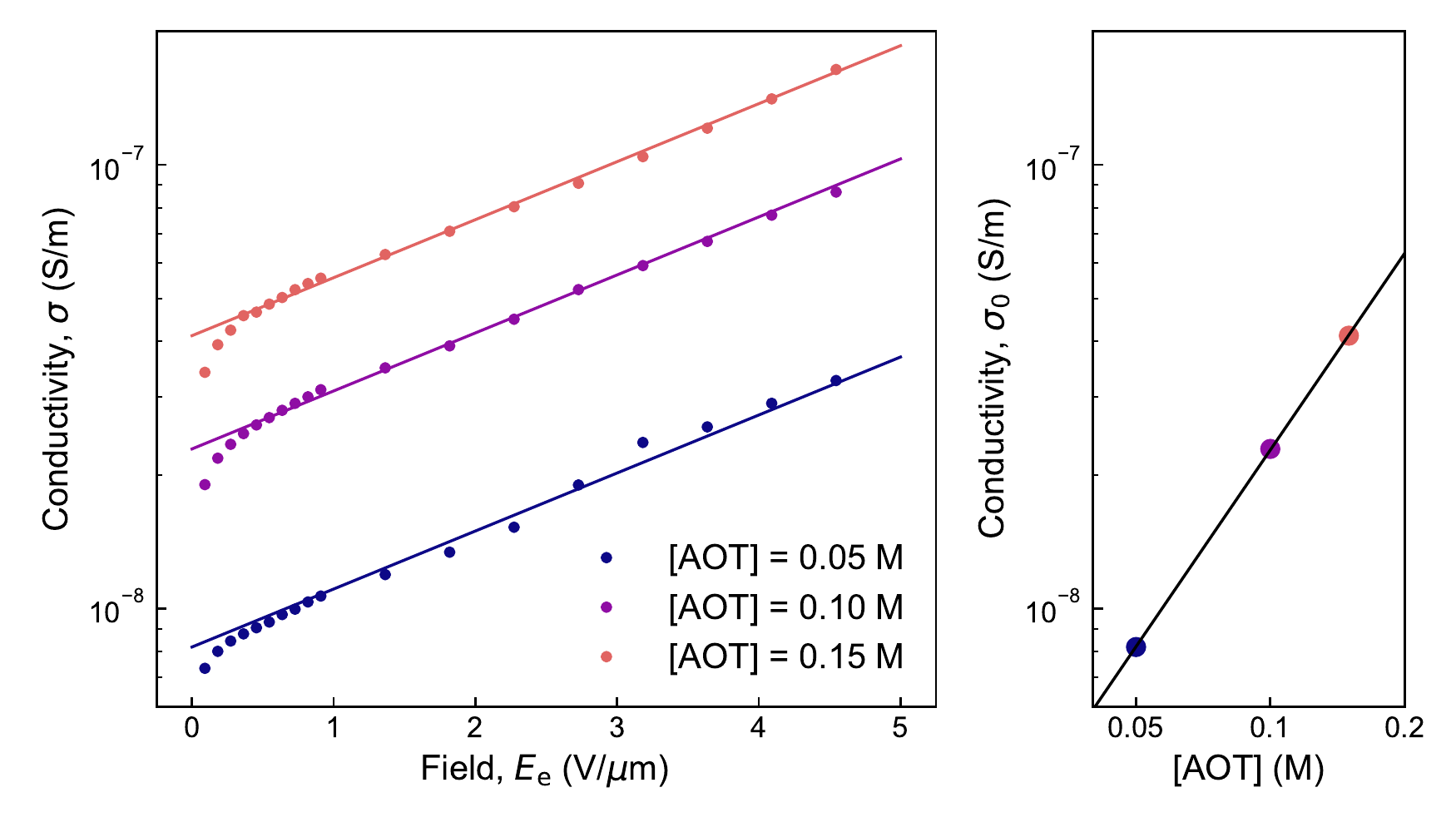}
    \caption{Conductivity $\sigma$ vs.\ field strength $E_{\e}$ for AOT-hexadecane solutions of different concentrations (left). The lines show best linear fits of the form $\sigma=\sigma_0 \exp(b E_{\e})$ with $b=0.301~\mu$m\,$V^{-1}$ and $\sigma_0=0.819$, $2.29$, and $4.12\times10^{-8}$ S\,m$^{-1}$ for AOT concentrations $\text{[AOT]}=0.05$, $0.1$, and $0.15$ M, respectively.  The zero field conductivity $\sigma_0$ increases faster than linearly with AOT concentration as approximated by the relation, $\sigma_0\propto \text{[AOT]}^{1.47}$ (right).}
    \label{fig:conductivity}
\end{figure}

We measured the conductivity of the AOT-hexadecane electrolytes using the same chamber used for the Quincke experiments described in the main text.  The electrode separation was $L=110~\mu$m; the electrode area was $A=1.0\times2.0$ cm$^2$. Using a sourcemeter (Keithley 2410), we applied different voltages $V$ across the chamber and measured the current $I$ and resistance $R=V/I$ at regular intervals of 0.15 s for a duration of six minutes. The conductivity $\sigma$ was evaluated from this data as 
\begin{equation}
    \sigma=\frac{\langle R\rangle A}{L} 
\end{equation}
where $\langle R\rangle$ denotes the average resistance over the duration of each experiment. Figure \ref{fig:conductivity} shows the conductivity $\sigma$ as a function of the applied field $E_{\e}=V/L$ for the three AOT concentrations used in our experiments.

\subsection{Zero-Field Conductivity}

The anionic surfactant, AOT (sodium bis-2-ethyl hexyl sulphosuccinate), forms inverse micelles in hexadecane above a  critical micelle concentration (CMC) of $\sim\!10^{-6}$ M.\autocite{Sainis2008}  Small-angle neutron scattering (SANS) data suggest that each inverse micelle contains $N\approx20$ AOT molecules with a polar core of radius $a_{\cc} \approx 1$ nm and a hydrodynamic radius of $a_{\h} \approx 1.7$ nm.\autocite{Kotlarchyk1985} At equilibrium,  the charge fluctuation theory\autocite{Prieve2017} predicts the number density of charged micelles to be
\begin{equation}
    n_z = n_0 \exp{\left(\frac{-z^2\lambda_{\B}}{2 a_{\cc}}\right)}
\end{equation}
where $n_0$ is the concentration of uncharged micelles, $z$ is the valence of charged micelles, and $\lambda_{\B}$ is the Bjerrum length
\begin{equation}
    \lambda_{\B} = \frac{e^2}{4\pi \varepsilon k_{\B} T} \approx 27 \text{ nm}
\end{equation}
where $e$ is the elementary charge, $\varepsilon = 2.05\varepsilon_0$ is the permittivity of hexadecane, $k_{\B}$ is the Boltzmann constant, and $T=298$ K is the temperature.  The fact that $\lambda_{\B}/2a_{\cc}\approx 14\gg 1$ implies that nearly all of the charged micelles have unit valence, $z=\pm 1$.  The equilibrium (zero-field) conductivity of the solution is estimated to be 
\begin{equation}
    \sigma_0 = \sum_z  z^2 e^2 \mu n_z\approx 2 e^2 \mu n_0 \exp{\left(-\frac{\lambda_{\B}}{2a_{\cc}}\right)}
\end{equation}
where $\mu\approx (6\pi\eta a_{\h})^{-1}$ is the micelle mobility, and $\eta=2.9$ mPa$\cdot$s is the dynamic viscosity of hexadecane.  

This model predicts that the conductivity increases linearly with the AOT concentration in agreement with previous experiments.\autocite{Sainis2008}  By contrast, our conductivity measurements at high AOT concentrations suggest that the conductivity increases faster than linearly with AOT concentration (Figure~\ref{fig:conductivity}, right).  This observation is explained by the increase in the dielectric constant of the solution with increasing AOT concentration. Using Maxwell's theory, the effective permittivity for a dispersion of particles (here, micelles) of permittivity $\varepsilon_{\p}$ in a fluid of permittivity $\varepsilon_{\f}$ is 
\begin{equation}
    \varepsilon = \varepsilon_{\f}\frac{2 \varepsilon_{\f} + \varepsilon_{\p} + 2 \phi (\varepsilon_{\p} - \varepsilon_{\p})}{2\varepsilon_{\f} + \varepsilon_{\p} - \phi (\varepsilon_{\p} - \varepsilon_{\f} )} = \varepsilon_{\f} + 3 \varepsilon_{\f}  \left(\frac{\varepsilon_{\p}-\varepsilon_{\f}}{\varepsilon_{\p} + 2 \varepsilon_{\f}}\right)\phi + O(\phi^2)
\end{equation}
where $\phi\approx \tfrac{4}{3}\pi a_{\h}^3 n_0$ is the volume fraction of particles (micelles).  The experimental data for the zero-field conductivity are well described by the parameter estimates, $a_{\cc}=1.16$ nm and $\varepsilon_{\p}=4.22\varepsilon_0$, which are physically reasonable. 

\subsection{Ionization Kinetics}

Two uncharged micelles can ``react'' to form two oppositely charged micelles with a rate approximated as 
\begin{equation}
    r = k_{\r}(n_{\eq}^2  - n_+ n_-)
\end{equation}
where $n_{\eq} = n_0 \exp(-\lambda_{\B}/2a_{\cc})$ is the equilibrium ion concentration predicted by the charge fluctuation theory, $n_{\pm}$ is the concentration of positive/negative ions, and $k_{\r}$ is the rate constant for ion-ion recombination.  As only a small fraction of micelles are charged, the concentration of neutral micelles can be approximated from the AOT concentration as $n_0=[\text{AOT}]/N$. If recombination is assumed to be diffusion-limited, the rate constant can be estimated\autocite{Debye1942,Saville1997} as 
\begin{equation}
    k_{\r} = 8\pi D \lambda_{\B} = \frac{2 e^2 \mu}{ \varepsilon} \approx 2.9\times 10^{-17} \text{ m}^3/\text{s} \label{eq:kr}
\end{equation}
where $D=k_{\B} T \mu$ is the micelle diffusivity.  The micelles approach their equilibrium ionization with a characteristic time scale $(2k_{\r} n_{\eq})^{-1}$, which is equal to one half of the charge relaxation time $\varepsilon/\sigma_0$. For a 0.1 M AOT solution, the equilibrium ion concentration is $n_{\eq}\approx 7\times 10^{-8}$ M and the ion relaxation time is $(2 k_{\r} n_{\eq})^{-1}\geq 0.4$ ms with equality for diffusion-limited recombination.

\subsection{Field-Dependent Conductivity}

The measured conductivity increases with the magnitude of the applied field as approximated by the empirical relation 
\begin{equation}
    \sigma = \sigma_0 \exp(b E_{\e})
\end{equation}
where $\sigma_0$ is the zero-field conductivity, and $b=0.301~\mu$m\,V$^{-1}$ is a fitted parameter independent of AOT concentration (Figure \ref{fig:conductivity}). This functional form and the value of $b$ can be rationalized by a simple kinetic model that describes the charging of two neutral micelles following their collision in the presence of the applied field. In this model, the rate of ionization is approximated as 
\begin{equation}
    r_{\ii} = k_{\ii} n_0^2 \exp\left(- \frac{U_{\aa}}{k_B T} \right)  \quad \text{with} \quad U_{\aa} = \frac{2e^2}{8\pi\varepsilon a_{\cc}} - \frac{e^2}{4\pi\varepsilon \lambda_{\B}} - e \lambda_{\B} E_{\e} 
\end{equation}
where $k_{\ii}$ describes the rate of collisions between neutral micelles present at concentration $n_0$. Here, $U_{\aa}$ describes the activation energy required to create a pair of oppositely charged micelles separated by the Bjerrum length $\lambda_{\B}$. It includes the Born self-energy of the two charged micelles, the Coulomb energy of interaction between them, and the energy of separating the charged pair in the electric field. Balancing the rate of ionization with that of recombination, $r_{\rr}=k_{\rr} n_+ n_-$, we obtain the following expression for the ion concentration
\begin{equation}
    n_{\pm} = n_0  \exp\left(-\frac{\lambda_{\B}}{2a_{\cc}}  +  b E_{\e} \right) \quad \text{with} \quad b = \frac{e \lambda_B}{2k_{\B} T} \approx 0.5~\mu\text{m\,V}^{-1}
\end{equation}
where we have chosen, $k_{\ii} = k_{\rr} e^{-1}$, to ensure consistency with the charge fluctuation theory in the zero field limit. The parameter $b$ agrees to within in a factor of two with the value determined experimentally.  Among the many simplifying approximations of this model, we have assumed that the process of ion-ion recombination does not depend on the magnitude of the external field.

\clearpage
\section{Data Collection \& Analysis}

\subsection{Experimental Details}
Polystyrene (PS) spheres of radius $a=1.5$, 2.5, or 5 $\mu$m (ThermoFisher Scientific) are washed and dried overnight at  55$^{\circ}$C prior to dispersing in hexadecane solutions of AOT surfactant (Sigma-Aldrich). Dilute particle suspensions (volume fraction $< 0.0002$) are injected into a microfluidic chamber created by two ITO-coated glass electrodes (Sigma-Aldrich) separated by one or more glass coverslips as spacers. Once filled, the chamber is sealed with epoxy. A constant voltage $V=$ 0--900 V is applied between the two electrodes by a source measure unit (Keithley 2410). The magnitude of resulting electric field $E_{\e}=V/L=$ 0--6 V/$\mu$m is limited primarily by the dielectric strength of the hexadecane solution.  The motion of the particles at the lower electrode is imaged from below using an optical microscope (Nikon Eclipse Ti-U) equipped with high speed camera (Phantom V310); videos are captured at 2000 frames per second. From the videos, we use TrackPy v0.4.2 in Python to detect particles and link their trajectories.  

\subsection{Bayesian Model Section}

For each track, we analyze the 2D particle position $\ve{x}=(x,y)$ as a function of time $t$ using three different models: stationary (denoted $M_s$), rolling ($M_r$), and oscillatory ($M_o$). We use Bayesian model selection to determine the posterior probability of each model given the tracking data. These probability assignments are used to color the markers in Figures 1c and 2 of the main text.

Each model is characterized by uncertain parameters denoted by the vector $\ve{w}$---for example, the initial particle position $\ve{x}_0$ and the particle velocity $\ve{U}$ in the rolling model. Given data consisting of the particle positions $\ve{x}_i$ at times $t_i$ (denoted $\mathcal{D}=\{\ve{x}_i,t_i\}$ for $i=1,...,n$), the posterior distribution for the parameters $\ve{w}$ of model $M$ is given by Bayes theorem 
\begin{equation}
    p(\ve{w}\mid\mathcal{D},M) = \frac{ p(\mathcal{D}\mid \ve{w},M) p(\ve{w}\mid M)}{ p(\mathcal{D}\mid M)} \label{eq:posterior}
\end{equation}
where $p(\ve{w}\mid M)$ is the prior, and $ p(\mathcal{D}\mid \ve{w},M)$ is the likelihood. The denominator, representing the evidence for the model $M$, is evaluated by marginalizing over the parameters $\ve{w}$ as 
\begin{equation}
    p(\mathcal{D}\mid M) = \int p(\ve{w}\mid \mathcal{D},M)p(\ve{w}\mid M) d\ve{w}
\end{equation}
This quantity is required in estimating the probability that model $M$ is true, relative to the competing models 
\begin{equation}
    p(M\mid \mathcal{D}) \propto p(\mathcal{D}\mid M) p(M)
\end{equation}
Here, we assume that each model has equal prior probabilities such that $p(M_s)=p(M_r)=p(M_o)$; the ``best'' model is that with the greatest evidence $p(\mathcal{D}\mid M)$.

\subsection{Stationary, Rolling, and Oscillatory Models}

For each experiment, we divide the reconstructed tracks for all particles into non-overlapping windows of $n$ frames ($n=20$ for $[\text{AOT}]=150$ mM, $n=60$ for $[\text{AOT}]=50$ and 100 mM); each window contains the positions $\{\ve{x}_i\}$ of a single particle at $n$ time points.  The size of the windows is long enough to distinguish between the three types of motion but short enough to neglect the changing orientation of particle rolling or oscillation. We analyze each window independently of the others and pool the resulting parameter estimates.  Within each window, the three models are given by 
\begin{align}
    M_s:\quad \ve{x}_i &= \ve{x}_0 + \xi_i
    \\
    M_r:\quad\ve{x}_i &= \ve{x}_0 + \ve{U} t_i + \xi_i
    \\
    M_o:\quad \ve{x}_i &= \ve{x}_0 + \ve{A}\cos(\omega t_i) + \ve{B}\sin(\omega t_i) + \xi_i
\end{align}
where $i=1,\dots,n$ and $\xi_i\sim\mathcal{N}(0,\sigma^2)$ is Gaussian noise with zero mean and standard deviation $\sigma = 1$ pixel (0.5 $\mu$m). The stationary and rolling models depend linearly on their parameters and are treated using standard methods for linear models (see below).

\subsubsection*{Normal Linear Models}

Here, we briefly summarize the necessary results for inference and model selection using normal linear models.\autocite{williams2006gaussian}  These results are applied in analyzing the experimental data using the stationary and rolling models, which have the following linear form
\begin{equation}
    \ve{y} = X^{\top} \ve{w} + \varepsilon 
\end{equation}
where $\ve{y}$ is an $n$-dimensional vector of outputs, $\ve{w}$ is a $D$-dimensional vector of weights (i.e., parameters of our linear model), $X$ is the $D\times n$ design matrix, and $\xi\sim\mathcal{N}(\ve{0},\sigma^2 I)$ is independent identically distributed Gaussian noise.  Given data consisting of the outputs $\ve{y}$ and the design matrix $X$, we aim to infer the values of the parameters $\ve{w}$ as 
\begin{equation}
    p(\ve{w}\mid \ve{y}, X) =  \frac{p(\ve{y}\mid \ve{w}, X)p(\ve{w})}{p(\ve{y}\mid X)}
\end{equation}
Here, the prior for the parameters $\ve{w}$ is assumed to be a normal distribution with zero mean and covariance matrix $\Sigma_p$
\begin{equation}
    p(\ve{w}) = \mathcal{N}(\ve{0},\Sigma_p) 
\end{equation}
The posterior distribution is also normally distributed 
\begin{align}
    p(\ve{y}\mid \ve{w}, X)p(\ve{w}) &= C \exp\left(-\frac{1}{2\sigma^2}(\ve{y}-X^{\top}\ve{w})^{\top}(\ve{y}-X^{\top}\ve{w})\right)\exp\left(-\frac{1}{2} \ve{w}^{\top} \Sigma_p^{-1} \ve{w}\right)
    \\
    & = C \exp\left(-\frac{1}{2}(\ve{w}-\bar{\ve{w}})^{\top} A^{-1} (\ve{w}-\bar{\ve{w}})\right) \exp\left(- \frac{\ve{y}^{\top}\ve{y}}{2\sigma^2} + \frac{1}{2} \bar{\ve{w}}^{\top} A^{-1}\bar{\ve{w}}\right)
\end{align}
where $\bar{\ve{w}}=A X\ve{y}/\sigma^2$ is the posterior mean, $A=(\Sigma_p^{-1}+X X^{\top}/\sigma^2)^{-1}$ is the posterior convariance matrix, and the constant $C=(2\pi \sigma^2)^{-n/2} |2\pi \Sigma_p|^{-1/2}$.  The evidence is 
\begin{align}
    p(\ve{y}\mid X) &= \int  p(\ve{y}\mid \ve{w}, X)p(\ve{w}) d\ve{w}
    \\
    &= \frac{|2\pi A|^{1/2}}{(2\pi \sigma^2)^{n/2} |2\pi \Sigma_p|^{1/2}} \exp\left(- \frac{\ve{y}^{\top}\ve{y}}{2\sigma^2} + \frac{1}{2} \bar{\ve{w}}^{\top} A^{-1}\bar{\ve{w}}\right)
\end{align}
The logarithm of the evidence provides a useful measure by which measure the performance of one model versus another
\begin{equation}
    \ln p(\ve{y}\mid X) = \frac{1}{2} \ln\frac{|A|}{|\Sigma_p|} - \frac{n}{2} \ln (2\pi \sigma^2) - \frac{\ve{y}^{\top}\ve{y}}{2\sigma^2} + \frac{1}{2} \bar{\ve{w}}^{\top} A^{-1}\bar{\ve{w}}
\end{equation}

\paragraph{Stationary Model.}
In the stationary model $M_s$, the unknown parameters $\ve{w}$ include only the constant particle position $\ve{x}_0$.  Prior to analysis, we center the tracking data for each window such that $\sum_i \ve{x}_i=0$.  We adopt a Gaussian prior for $\ve{x}_0$ with zero mean and covariance matrix $\Sigma_p = I(10 \text{ pixels})^2$, where $I$ is the $2\times2$ identity matrix. We use the same prior on $\ve{x}_0$ in the rolling and oscillatory models.

\begin{figure}[t]
    \centering
    \includegraphics[width=\textwidth]{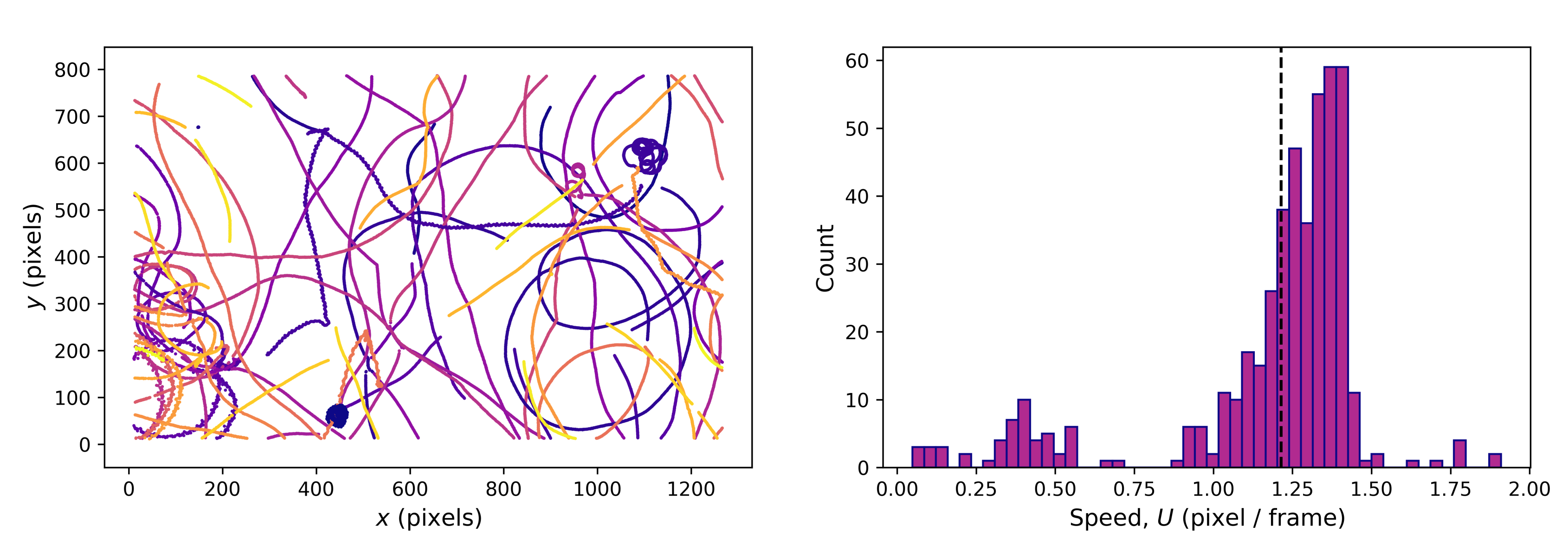}
    \caption{(left) Tracking data for 10 $\mu$m spheres immersed in a 0.15 M AOT hexadecane solution between parallel electrodes separated by 150 $\mu$m and subject to a voltage of 525 V. The tracking data was divided into 636 windows of $n=50$ frames, of which 73\% were classified as rolling.  (right) Histogram of the MAP estimates for particle speed $U$ for those 462 windows classified as rolling.  The dashed line shows the mean speed.}
    \label{fig:rolling}
\end{figure}

\paragraph{Rolling Model.}
In the rolling model, the unknown parameters $\ve{w}$ are augmented to include the particle velocity $\ve{U}$ as well as the initial position $\ve{x}_0$.  We adopt a Guassian prior for the velocity with zero mean and a standard deviation of 10 pixels/frame (ca.\ 5,000 $\mu$m/s) in both the $x$ and $y$-directions.  Figure \ref{fig:rolling} shows a typical result for the speeds $U$ inferred from windows classified as ``rolling''.  

\paragraph{Oscillatory Model.}
The oscillatory model is non-linear in the frequency $\omega$ but linear in the other parameters. Building on our review of linear models in the previous section, we now have a model of the form
\begin{equation}
    \ve{y} = X^{\top}_{\omega} \ve{w} + \varepsilon \label{eq:oscillatory}
\end{equation}
where the design matrix depends non-linearly on the frequency $\omega$. For a given value of the frequency, this model can be analyzed using the methods of the previous section. The posterior for the frequency can be written as 
\begin{equation}
    p(\omega \mid \ve{y}) = \frac{p(\ve{y}\mid X_{\omega})p(\omega)}{p(\ve{y})}
\end{equation}
We use a uniform prior on the domain $2\pi/n\Delta t < \omega < \pi/\Delta t$, where $n$ is the number of frames in the window, and $\Delta t$ is the time step between consecutive frames  (i.e., $\Delta t = 1$ frame). The minimum frequency corresponds to a single oscillation cycle over the length of the window; the maximum frequency is the Nquist frequency. We use numerical optimization to identify the maximum a posterior probability (MAP) estimate of the frequency $\omega^{*}$. We use this point estimate of the frequency to analyze the oscillatory model (\ref{eq:oscillatory}) as a linear model using the methods of the previous section.

\paragraph{Gaussian Process Model.}
For some windows, none of the three models describe the data well. We therefore introduce a forth model using a Gaussian process\autocite{williams2006gaussian}, which provides a general purpose alternative.  The Gaussian process models the data as 
\begin{equation}
    y = f(t) + \varepsilon \quad\text{with}\quad f(t) = \mathcal{GP}(m(t),k(t,t'))
\end{equation}
Here, the mean function is assumed to be zero, $m(t)=0$, and the covariance function (or kernel) is chosen as a squared exponential
\begin{equation}
    k(t,t') = \sigma_f^2 \exp(\tfrac{1}{2}|t-t'|^2 / \tau)
\end{equation}
with hyperparameters $\sigma_f$ and $\tau$ characterizing the amplitude and correlation time, respectively.  We use values of $\sigma_f=5$ pixels and $\tau=2$ frames for the analysis of all experiments. Gaussian process regression is implemented using the scikit-learn package in Python, which provides the log-evidence $\ln p(\ve{y}\mid \sigma_f,\tau)$ needed for model selection.

\clearpage
\section{Additional Experimental Data}

\subsection{Effect of AOT Concentration $[\text{AOT}]$}
\vspace{1cm}
\begin{figure}[h]
    \centering
    \includegraphics{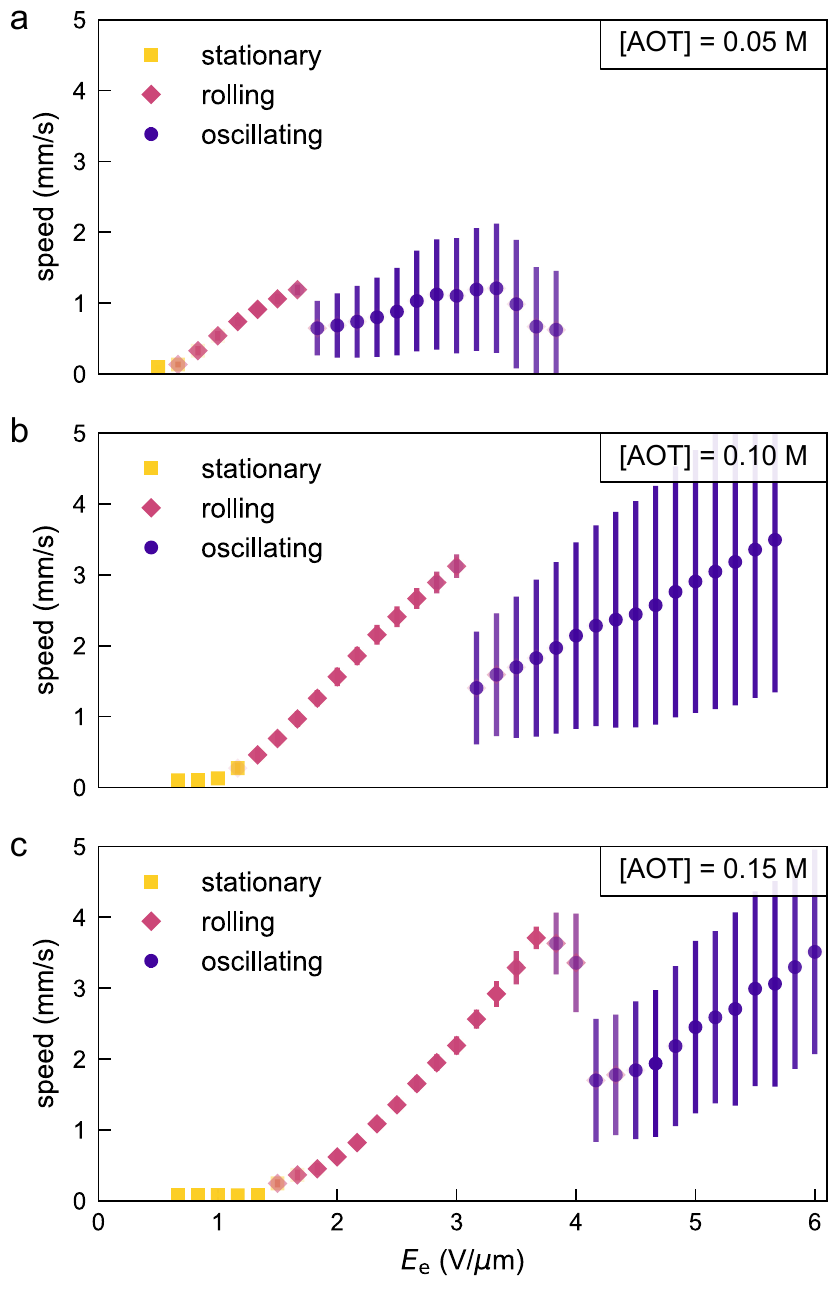}
    \caption{Time-averaged particle speed vs.\ external field strength $E_{\e}$ for $a=5~\mu$m PS spheres at three different AOT concentrations. The electrode separation is $L=150~\mu$m. The plot is (c) is reproduced from Figure 1c in the main text. As in Figure 1c, we compute the mean and standard deviation of the particle speed for each 20 ms trajectory. Markers denote the median of these mean speeds; error bars denote the median of the corresponding standard deviations. The plotted data are colored based on probability assignments of the Bayesian classifier.}
    \label{fig:aot_concentration}
\end{figure}

\clearpage
\subsection{Effect of Particle Radius $a$}
\vspace{1cm}
\begin{figure}[h]
    \centering
    \includegraphics{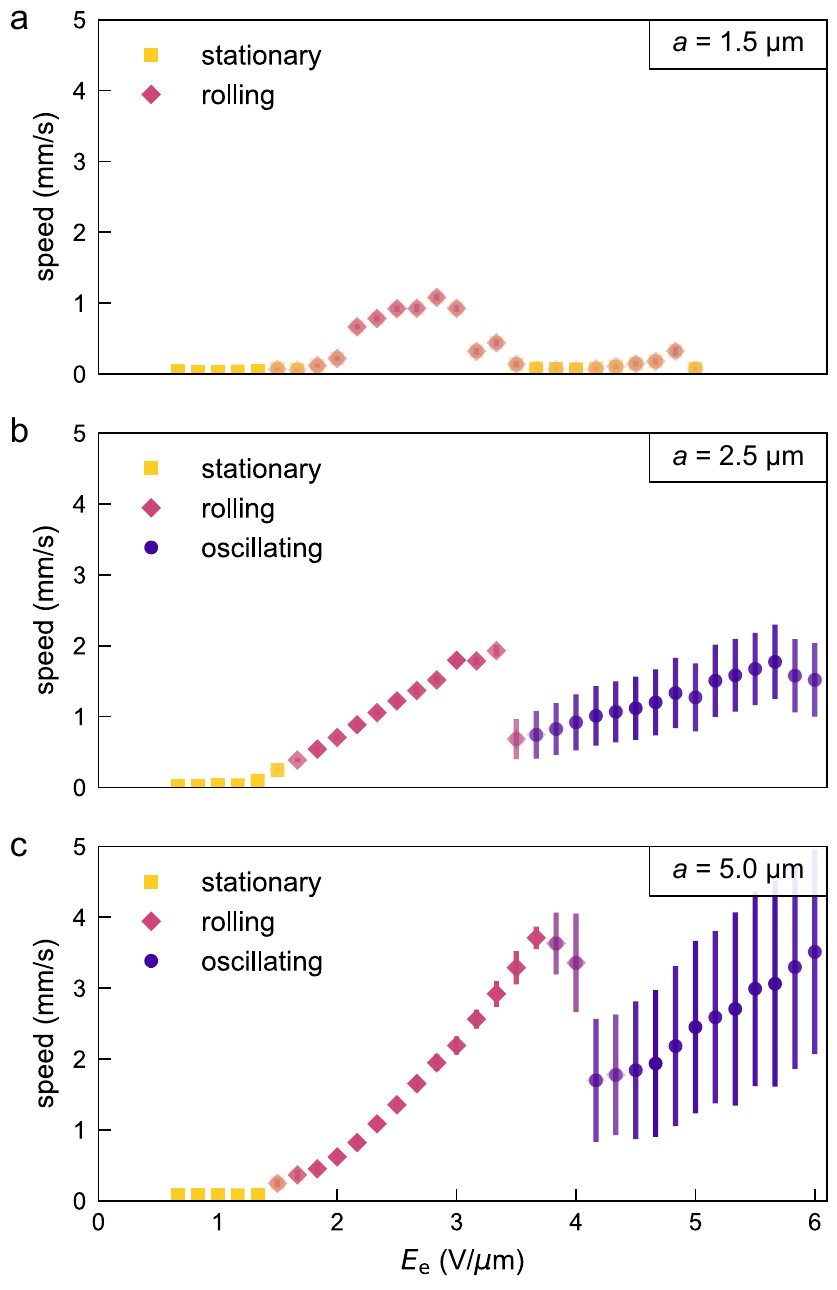}
    \caption{Time-averaged particle speed vs.\ external field strength $E_{\e}$ for PS spheres of three different radii. The AOT concentration is $\text{[AOT]}=0.15$ M; electrode separation is $L=150~\mu$m. The plot is (c) is reproduced from Figure 1c in the main text. As in Figure 1c, we compute the mean and standard deviation of the particle speed for each 20 ms trajectory. Markers denote the median of these mean speeds; error bars denote the median of the corresponding standard deviations. The plotted data are colored based on probability assignments of the Bayesian classifier.}
    \label{fig:particle_size}
\end{figure}

\clearpage
\subsection{Oscillatory Rolling in the Transition Region}

\vspace{1cm}

\begin{figure}[h]
   \centering
    \includegraphics{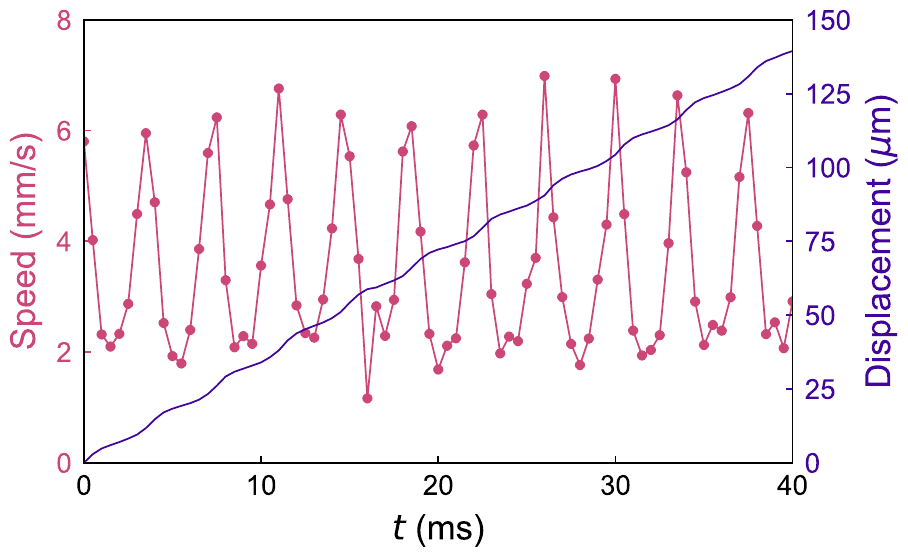}
    \caption{Instantaneous speed and accumulated displacement of a PS particle with radius $a=5~\mu$m as a function of time for a field strength $E_{\e}=4~$V/$\mu$m. The AOT concentration is $\text{[AOT]}=0.15$ M; electrode separation is $L=150~\mu$m. These conditions correspond to the transition between steady rolling and back-and-forth oscillations (see Figure 1c).}
    \label{fig:transition}
\end{figure}


\clearpage
\subsection{Arrested Motion of $1~\mu$m Particles}
\vspace{1cm}
\label{sec:1um}
\begin{figure}[htp]
   \centering
    \includegraphics{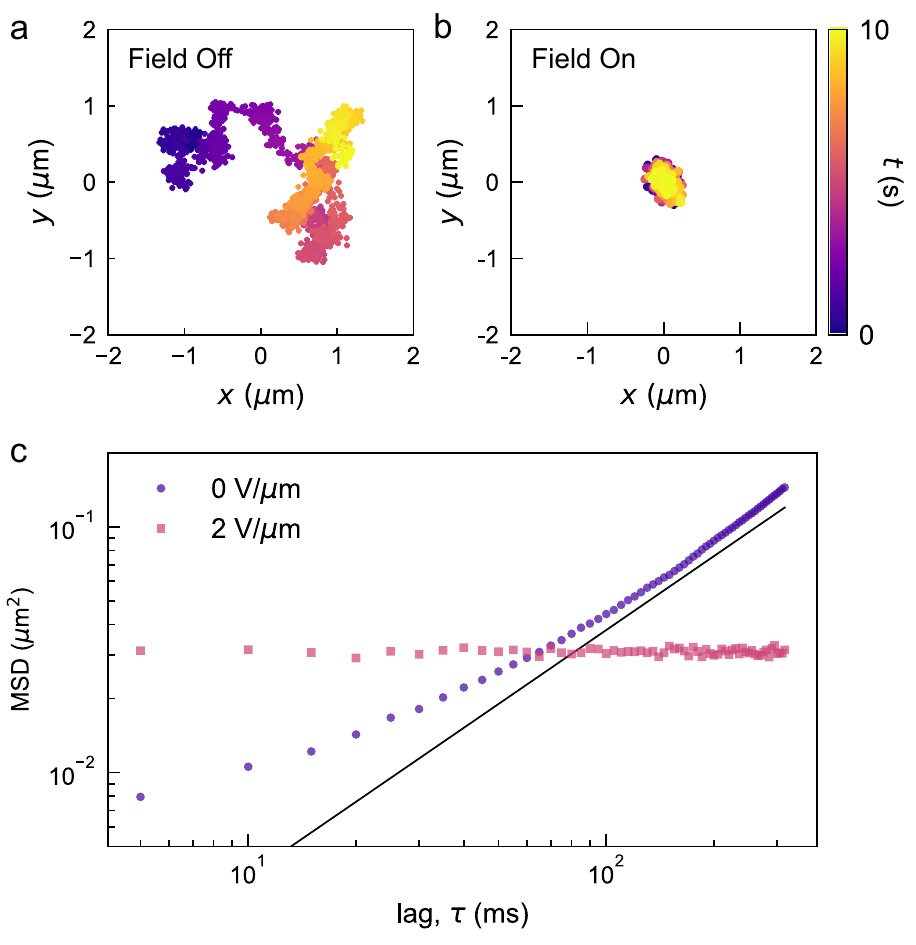}
    \caption{(a) Brownian trajectory of a 1 $\mu$m PS sphere in a 150 mM solution of AOT in hexadecane. (b) Application of an external field $E_{\e}=2$ V/$\mu$m arrests the particle's motion. (c) Mean squared displacement (MSD) vs.\ lag time for the trajectories in (a) and (b). In the absence of the field, the particle has an apparent diffusivity of $D_{\p}=0.095~\mu\text{m}^2$/s. The particle does not diffuse upon application of the field.}
\end{figure}

\clearpage
\subsection{Oscillation Frequency $\omega \tau_{\mw}$ vs.~Field Strength $E_{\e}/E_{\cc}$ }
\vspace{1cm}
\begin{figure}[h]
   \centering
    \includegraphics{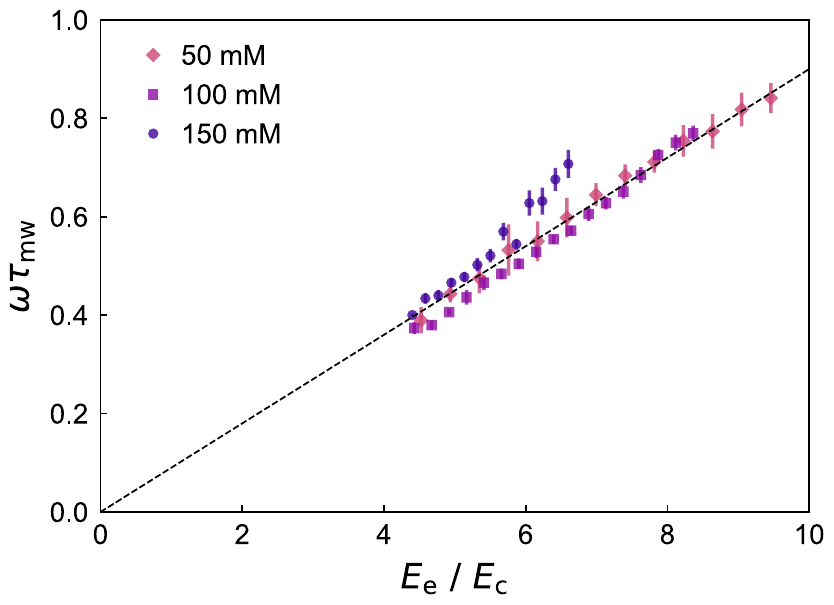}
    \caption{Oscillation frequency $\omega$ scaled by the Maxwell-Wager time $\tau_{\mw}$ vs.\ external field strength $E_{\e}$ scaled by the critical field strength $E_{\cc}$ for three different AOT concentrations. These data are identical to those shown in Figure 3b but are here scaled by $\tau_{\mw} = 3.5, 1.3, 0.70$ ms and $E_{\cc}=0.41, 0.68, 0.91$ V/$\mu$m for $[\text{AOT}]=50$, 100, 150 mM, respectively. The particle radius is $a=5~\mu$m.  Markers denote the mean frequencies within populations of particle trajectories of equal duration; error bars denote standard deviations of these populations. The dashed curve is the equation $\omega \tau_{\mw} = 0.09 E_{\e} / E_{\cc}$ stated in the main text.}
\end{figure}

\clearpage
\subsection{Acoustic Levitation Experiments}
\vspace{1cm}
\begin{figure}[h]
   \centering
    \includegraphics{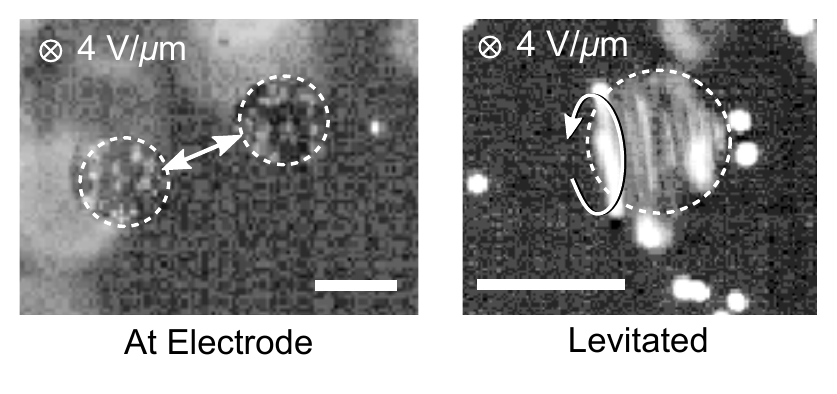}
    \caption{Time-lapse microscopy images showing the dynamics of a single particle oscillating at the electrode surface (left) and rotating within the bulk fluid (right). Dashed circles highlight the particle position. The external field strength in both experiments is $E_{\e}=4~$V/$\mu$m. Scale bars are $40~\mu$m. Here, we use 40 $\mu$m polyethylene Janus spheres with a white  hemisphere coated on black core (HCMS-BLK-WHT-Polymer from Cospheric Inc.).  Also visible in the images are 1 $\mu$m fluorescent particles (white), which aid in visualizing the motion of the larger particle. Particle levitation is achieved using a piezoelectric transducer adhered to the outside of the top electrode. The transducer is driven by a sinusoidal voltage signal with an amplitude of 10 V and a frequency of 3.6 MHz. The precise frequency is tuned to create a standing acoustic field with wavelength equal to twice the electrode separation $L=200~\mu$m. Under these conditions, particles levitate at the mid-plane equidistant from the two parallel electrodes.}
    \label{fig:levitated}
\end{figure}

\clearpage
\section{Analysis of a Model Electrolyte in 1D}
\label{sec:model_electrolyte}

Following Saville,\autocite{Saville1997} we consider a binary monovalent electrolyte, in which the charge carriers are created by the homogeneous ionization of neutral species.  The concentration of positive and negative charge carriers is governed by the Nernst-Planck equation
\begin{equation}
    \frac{\partial n_\pm}{\partial t}+\nabla\cdot \left(n_{\pm}\ve u  -D \nabla n_\pm \pm \frac{e D}{k_{\B} T} n_\pm \ve E \right) = k_{\r} (n_{\eq}^2 - n_+n_-) \label{eq:npm}
\end{equation}
where $\ve u$ is the fluid velocity. The electric field $\ve{E}=-\nabla \phi$ is related to the local charge density by the Poisson equation
\begin{equation}
    \nabla^2 \phi = -\frac{e}{\varepsilon}(n_+ - n_-) \label{eq:phi}
\end{equation}
Here, we analyze the behavior of this model---the so-called Poisson-Nernst-Planck (PNP) equations---in 1-dimension for different boundary conditions in the presence of strong electric fields $E\gg k_{\B} T/e \ell$. First, we consider a semi-infinite electrolyte in contact with an absorbing boundary at which the concentration of charge carriers is zero. Analysis of this problem reveals the electric boundary layer of thickness $\ell$ anticipated by the scaling arguments presented in the main text. Next, we consider a 1D electrolyte confined between two absorbing boundaries.  We show how the electric current is limited by the rate of ionization when the domain size is smaller than the boundary layer thickness.  Finally, we consider the effects of electrolyte confinement on the transient charging of a no-flux boundary.   

\subsection{Semi-Infinite Electrolyte with an Absorbing Boundary}

We first consider an absorbing boundary at $x=0$ in contact with our model electrolyte on the domain $x>0$.  Introducing the total ion concentration, $n_{\s}=n_+ + n_-$, and the concentration difference, $n_{\dd}=n_+ - n_-$, the governing equations can be written as
\begin{align}
\begin{split}
    \frac{\partial n_{\dd}}{\partial t} + e \mu \frac{\partial(n_{\s} E)}{\partial x}&= 0
    \\
    \frac{\partial n_{\s}}{\partial t} + e \mu \frac{\partial(n_{\dd} E)}{\partial x} &= 2k_{\r} \left[n_{\eq}^2-\frac{1}{4}\left(n_{\s}^2-n_{\dd}^2\right)\right]
    \\
    \frac{\partial E}{\partial x} &= \frac{e}{\varepsilon} n_{\dd} \label{eq:1D}
\end{split}
\end{align}
where $\mu = D/k_{\B} T\approx (6\pi \eta a_{\h})^{-1}$ is the mobility of the charged micelles. Here, we have omitted convective and diffusive transport which are negligible for the strong fields of interest as confirmed below. Due to our neglect of the diffusive flux, it is not possible to simultaneously satisfy the conditions, $n_{+}(0)=0$ and $n_{-}(0)=0$, at the absorbing boundary.  For a field in the positive $x$-direction ($E_{\e}>0$), we require only that the concentration of positive ions vanish at the boundary 
\begin{equation}
    n_+(0) = \frac{1}{2} \left(n_{\s}(0) + n_{\dd}(0)\right) = 0 \label{eq:1D_bc1}
\end{equation}
By contrast, the concentration of negative ions is expected to fall to zero within a thin diffusive boundary layer neglected in our present analysis. Far from the boundary ($x\rightarrow\infty$), the ion concentrations approach their equilibrium values in the presence of an external field $E_{\e}$
\begin{equation}
    n_{\dd}(\infty) = 0, \quad n_{\s}(\infty) = 2 n_{\eq},  \quad E(\infty)=E_{\e} \label{eq:1D_bc2}
\end{equation}

At steady-state, equations (\ref{eq:1D}) can be combined to give a single differential equation for the field squared $E^2$. 
Charge conservation implies that the electric current density is everywhere constant and equal to $j_{\e}= e^2 \mu n_{\s} E = 2 e^2 \mu n_{\eq} E_{\e}$, where the second equality follows from the boundary condition (\ref{eq:1D_bc2}).  The total ion concentration is therefore related to the applied field as $n_{\s} =2 n_{\eq} E_{\e}/E$. The concentration difference $n_{\dd}$ is related to the field by Gauss's law. Substituting these expressions for $n_{\dd}$ and $n_{\s}$ into the second line of equation (\ref{eq:1D}), we obtain the following nonlinear equation for the field squared
\begin{equation}
    \frac{d^2 E^2}{d x^2} = \frac{4 k_{\r} n_{\eq}^2}{\mu \varepsilon} \left[1 - \frac{E_{\e}^2}{E^2} + \frac{1}{E^2} \left(\frac{\varepsilon}{4e n_{\eq}} \frac{d E^2}{d x} \right)^2 \right] \label{eq:E2}
\end{equation}
At $x=0$, the boundary condition (\ref{eq:1D_bc1}) becomes
\begin{equation}
    2 n_{\eq} \frac{E_{\e}}{E(0)} + \frac{\varepsilon}{e} \left.\frac{d E}{dx}\right\rvert_0 = 0 \quad \text{such that} \quad \left.\frac{d E^2}{d x}\right|_0 = -\frac{4 en_{\eq} E_{\e}}{\varepsilon}  \label{eq:E2bc1}
\end{equation}
Far from the boundary, the field approaches the asymptotic value $E(x\rightarrow \infty) = E_{\e}$.  

Equation (\ref{eq:E2}) is characterized by a length scale $\lambda$ over which the field approaches its asymptotic value.  Linearizing about this value, perturbations in the field squared---denoted $u = E^2 - E_{\e}^2$---are governed by 
\begin{equation}
    \frac{d^2 u}{d x^2} = \frac{4 k_{\r} n_{\eq}^2}{\mu \varepsilon E_{\e}^2} u + O(u^2)
\end{equation}
Perturbations decay over a characteristic length scale 
\begin{equation}
    \lambda = \frac{E_{\e}}{2 n_{\eq}} \sqrt{\frac{\mu \varepsilon}{k_{\r}}} = \frac{1}{2} \beta \ell
\end{equation}
This length scale $\lambda$ is closely related to the scale, $\ell = e\mu E_{\e}/k_{\r} n_{\eq}$, described in the introduction of the main text. The two are related by the dimensionless parameter $\beta =(k_{\r} \varepsilon / e^2 \mu)^{1/2}$, which characterizes the relative rates of charge formation ($2 k_{\r} n_{\eq}$) and charge relaxation ($2 e^2\mu n_{\eq}/\varepsilon$). Assuming the diffusion-limited approximation (\ref{eq:kr}) for the recombination rate constant $k_{\r}$, these rates are approximately equal, $\beta=\sqrt{2}$, as are the two length scales, $\ell/\lambda =\sqrt{2}$. For a 0.1 M AOT-hexadecane solution subject to an external field of $E_{\e}= 5$ V/$\mu$m, this length is estimated to be $\lambda\approx 5~\mu$m.  For comparison, the Debye length is $\lambda_{\DD}=(\varepsilon k_{\B} T/2e^2 n_{\eq})^{1/2}\approx 0.2~\mu$m.

\begin{figure}[t]
    \centering
    \includegraphics[width=9cm]{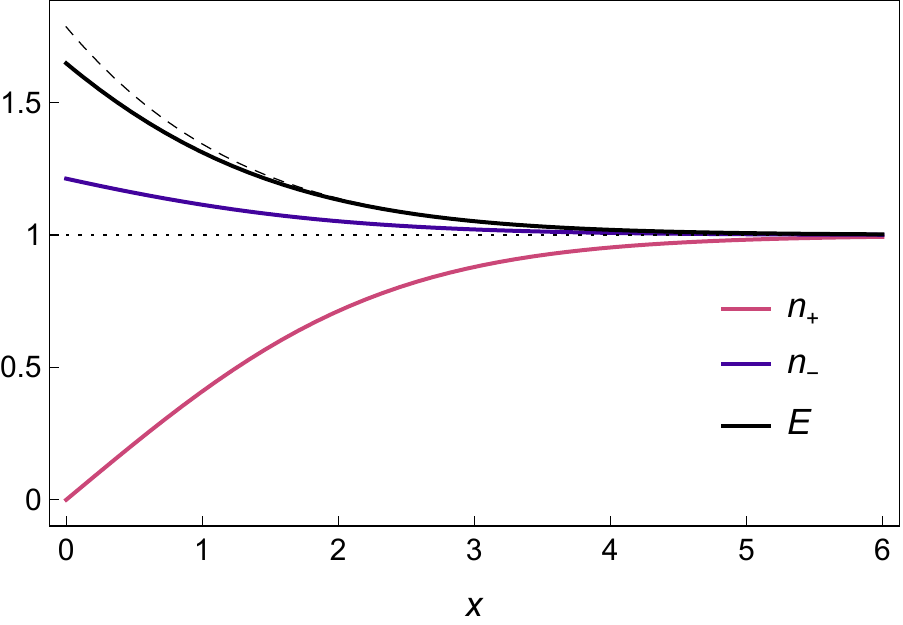}
    \caption{\textbf{One Absorbing Boundary.} Number density of positive and negative ions, $n_+$ and $n_-$, and the electric field $E$ as a function of distance $x$ from a plane boundary. Here, ion concentrations are scaled by the equilibrium concentration $n_{\eq}$, the field by the external field $E_{\e}$, and position by the characteristic length $\lambda$. The recombination parameter is $\beta^2=2$ corresponding to diffusion-limited ion recombination. The dashed green curve shows the asymptotic prediction of the linearized model.}
    \label{fig:Profiles}
\end{figure}

\paragraph{Numerical Solution.} Figure \ref{fig:Profiles} illustrates the electric field and the ion distribution computed numerically for the experimental conditions.  To facilitate our analysis, we non-dimensionalize the problem using the following scales
\begin{align}
\begin{split}
    \text{Concentration:} &\quad n_{\eq}
    \\
    \text{Field:} &\quad E_{\e}
    \\
    \text{Length:} &\quad \lambda = \frac{E_{\e}}{2 n_{\eq}} \sqrt{\frac{\mu \varepsilon}{k_{\r}}}
    \\
    \text{Time:} &\quad \frac{\varepsilon}{2 e^2 \mu n_{\eq}} 
    \label{eq:E2_scales}
\end{split}
\end{align}
The dimensionless field is governed by
\begin{equation}
    \frac{d^2 E^2}{d x^2} = 1 - \frac{1}{E^2} + \frac{\beta^2}{4 E^2} \left( \frac{d E^2}{d x} \right)^2  \label{eq:E2_nd}
\end{equation}
where $\beta^2 = k_{\r} \varepsilon / e^2\mu$ is the ratio between the rate of charge formation ($2 k_{\r} n_{\eq}$) and that of charge relaxation ($2 e^2\mu n_{\eq}/\varepsilon$). For simplicity of notation, we used the same symbols to denote dimensional and dimensionless quantities alike; the distinction should be made clear by the context. At the boundary ($x=0$), the condition that the positive ion concentration goes to zero implies that 
\begin{equation}
    \left.\frac{d E^2}{d x}\right|_0 = -\frac{2}{\beta}  \label{eq:E2_nd_bc1}
\end{equation}
Far from the boundary, the solution to the linearized equation implies that 
\begin{equation}
    \frac{d E^2}{d x} = 1 - E^2 \quad \text{for} \quad x \gg 1 \label{eq:E2_nd_bc2}
\end{equation}
We compute $E^2$ numerically on a finite domain using equation (\ref{eq:E2_nd}) subject to the boundary conditions (\ref{eq:E2_nd_bc1}) and (\ref{eq:E2_nd_bc2}).  Given the field, the dimensionless ion concentrations are computed as 
\begin{equation}
    n_{\s} = \frac{2}{E},  \quad n_{\dd} = 2 \beta \frac{d E}{d x}
\end{equation}

\paragraph{Validating Assumptions.}  In deriving the above results, we neglected the convective and diffusive contribution to the ion flux.  To validate these assumptions, we compare the magnitude of the terms retained to those neglected.  Diffusive contributions to the flux are small when
\begin{equation}
    \frac{D n_{\eq}}{\lambda} \ll \frac{e D}{k_{\B} T} n_{\eq} E_{\e} \quad \text{such that} \quad \frac{ e \lambda E_{\e} }{k_{\B} T} \gg 1 
\end{equation}
For a 0.1 M AOT-hexadecane solution subject to an external field of $E_{\e} = 5$ V/$\mu$m, this condition is satisfied as $e \lambda E_{\e}/k_B T\sim 10^3\gg $1.  Diffusive transport becomes relevant near the boundary within a thin region of thickness $k_{\B} T/e E_{e}\sim 10$ nm.   Similarly, convective transport is negligible provided that the characteristic flow velocity $U$ is sufficiently small
\begin{equation}
    U \ll e \mu E_{\e} \sim 10 \text{ mm/s}
\end{equation}

\begin{figure}[t]
    \centering
    \begin{subfigure}{.5\textwidth}
        \centering
        \includegraphics[width=0.95\textwidth]{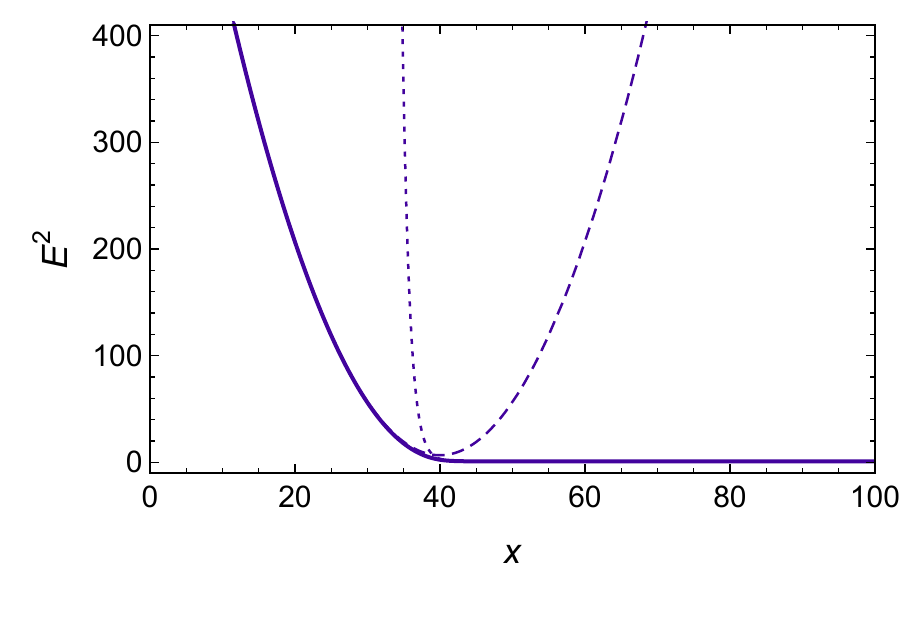}
    \end{subfigure}%
    \begin{subfigure}{.5\textwidth}
        \centering
        \includegraphics[width=0.95\textwidth]{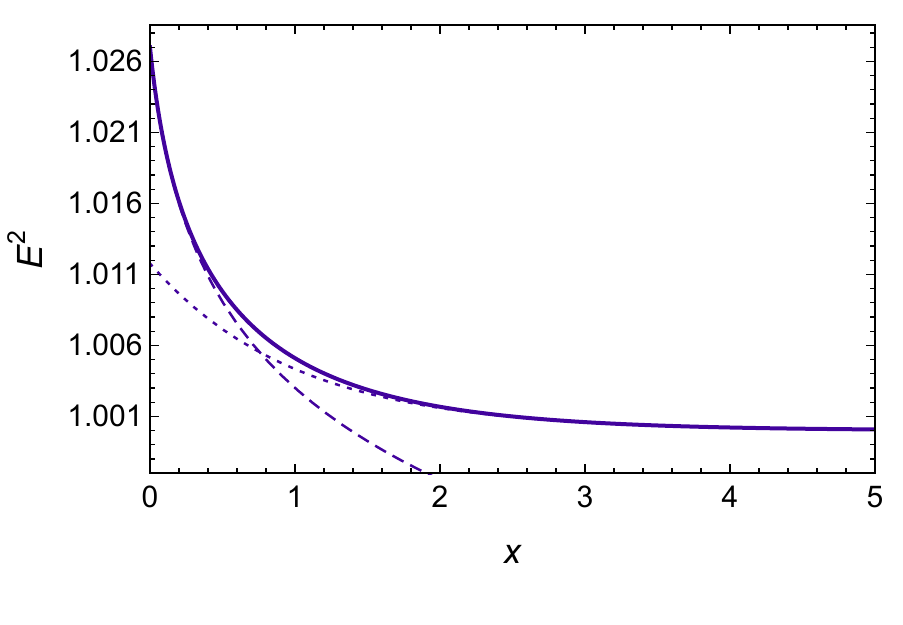}
    \end{subfigure}
    \caption{\textbf{Limiting Regimes.} Field squared $E^2$ vs.~position $x$ for slow recombination with $\beta = 0.05$  (left) and fast recombination with $\beta=20$ (right). Solid curves denote the numerical solution of equation (\ref{eq:E2_nd}) with boundary conditions (\ref{eq:E2_nd_bc1}) and (\ref{eq:E2_nd_bc2}). The field is scaled by the external field $E_{\e}$, and position by the characteristic length $\lambda$. For $\beta\ll1$, dashed and dotted curves denote the approximate relations, $ E^2\approx x^2/2 - 2x/\beta + C_1$ and $E^2\approx 1 - C_2 e^{-(x-2/\beta)}$, where the constants $C_1\sim 2\beta^{-2}$ and $C_2\sim 2$ are determined by matching these two solutions at $x\sim 2\beta^{-1}$. For $\beta\gg1$, dashed and dotted curves denote the approximate relations, $ E^2 \approx -4\beta^{-2}\ln \left(\beta/2 +\beta ^2 x/4 \right) + C_1 $  and $E^2 \approx 1 - C_2 e^{-(x-1)} $, where the constants $C_1\sim 1$ and $C_2\sim 4 \beta^{-2}$ are determined by matching these two solutions at $x\sim 1$. }
    \label{fig:limiting}
\end{figure}

\paragraph{Limiting Regimes.} The results above describe the specific scenario of diffusion-limited recombination, for which $\beta^2 = 2$. Here, we consider the structure of the solution for the limiting scenarios of $\beta\ll1$ and $\beta\gg1$. When the rate of ion recombination is much slower than the diffusion-limited rate ($\beta\ll1$; Figure \ref{fig:limiting}, left), the solution can be divided in two regions characterized by the following dominant terms in the dimensionless governing equation (\ref{eq:E2_nd})
\begin{align}
    \text{Region 1: } \quad \frac{d^2 E^2}{d x^2} &\approx 1 \quad \text{for} \quad  0\leq x \ll \frac{2}{\beta} 
    \\
    \text{Region 2: } \quad \frac{d^2 E^2}{d x^2} &\approx E^2 - 1 \quad \text{for} \quad  x \gg\frac{2}{\beta} 
\end{align}
Under these conditions ($\beta\ll1$), the thickness of the boundary region becomes $\ell = 2\lambda/\beta$ as predicted by the simple scaling argument in the main text.

The opposite limit of $\beta\gg1$ is unphysical as it implies ion-ion recombination faster than the diffusion-limited rate. We consider it here for the sake of completeness (Figure \ref{fig:limiting}, right). As above, the solution is divided in two regions characterized by the following dominant balances
\begin{align}
    \text{Region 1: } \quad \frac{d^2 E^2}{d x^2} &\approx \frac{\beta^2}{4}\left(\frac{d E^2}{d x}\right)^2 \quad \text{for} \quad  0\leq x \ll 1 
    \\
    \text{Region 2: } \quad \frac{d^2 E^2}{d x^2} &\approx E^2 - 1 \quad \text{for} \quad  x \gg 1
\end{align}

\subsection{Finite Electrolyte between Two Absorbing Boundaries}

\begin{figure}[b!]
    \centering
    \begin{subfigure}{.5\textwidth}
        \centering
        \includegraphics[width=0.95\textwidth]{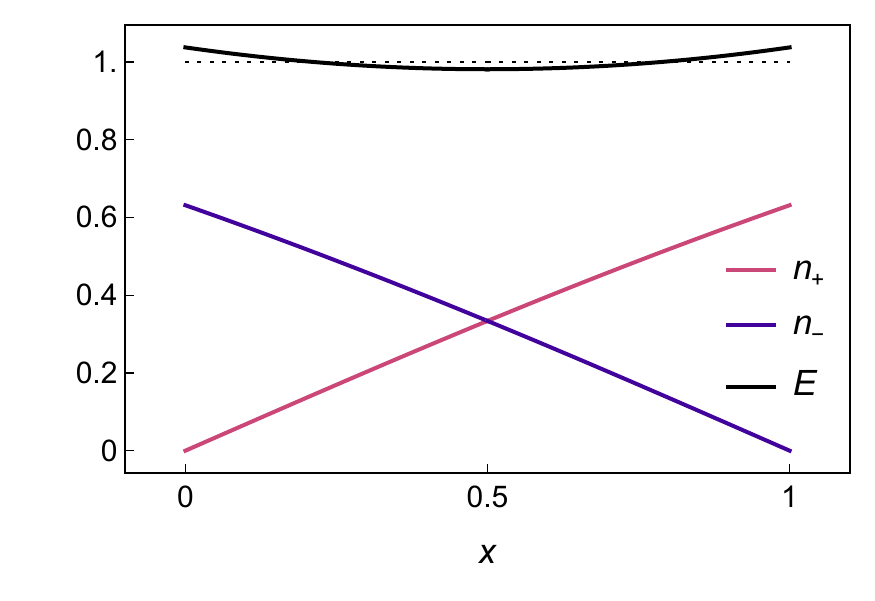}
    \end{subfigure}%
    \begin{subfigure}{.5\textwidth}
        \centering
        \includegraphics[width=0.95\textwidth]{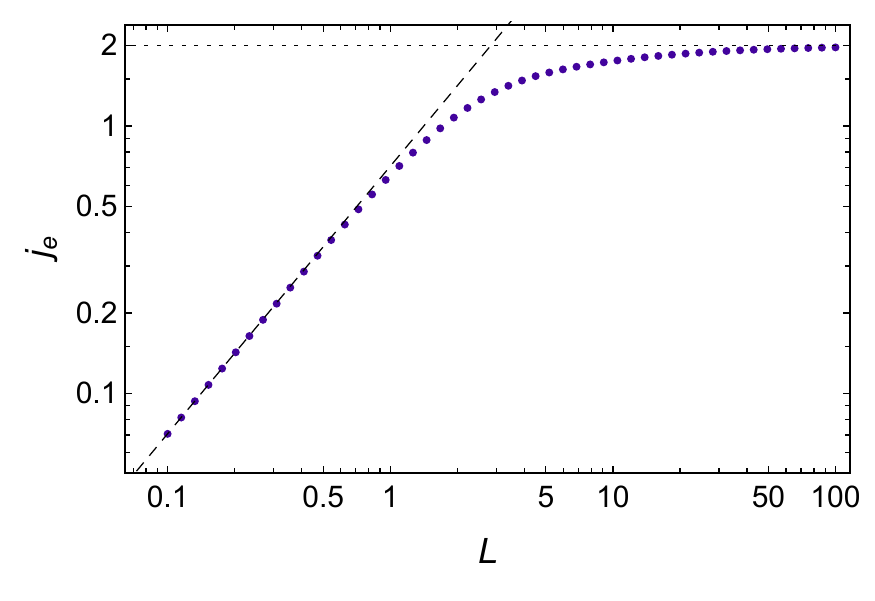}
    \end{subfigure}
    \caption{\textbf{Two Absorbing Boundaries.} (left) Number density of positive and negative ions, $n_+$ and $n_-$, and the electric field $E$ as a function of position $x$ between two absorbing boundaries separated by $L=1$. Ion concentrations are scaled by the equilibrium concentration $n_{\eq}$, the field by the external field $E_{\e}$, and position by the characteristic length $\lambda$. The recombination parameter is $\beta^2=2$ corresponding to diffusion-limited ion recombination. (right) Electric current density $j_{\e}$ (scaled by $e^2 \mu n_{\eq} E_{\e}$) as a function of the boundary separation $L$.  For small separations ($L\ll2$), the current is $j_{\e}\approx \beta L/2$; for large separations ($L\gg2$), it approaches $j_{\e}=2$.}
    \label{fig:Current}
\end{figure}

We now consider our model electrolyte sandwiched between two absorbing boundaries separated by a distance $L$ (Figure \ref{fig:Current}).  The potential difference across the electrolyte is specified as $L E_{\e}$ such that 
\begin{equation}
    \int_0^L E(x) d x = L E_{\e}
\end{equation}
This condition is used to determine the electric current density $j_{\e}$, which is now unknown.  Using the same scales (\ref{eq:E2_scales}), the dimensionless governing equation for the field squared is 
\begin{equation}
    \frac{d^2 E^2}{d x^2} =1 - \frac{j_{\e}^2}{4 E^2} + \frac{\beta^2}{4 E^2} \left( \frac{d E^2}{d x} \right)^2 
\end{equation}
where the current $j_{\e}$ is scaled by $e^2 \mu n_{\eq} E_{\e}$.  The conditions at the absorbing boundaries are
\begin{equation}
    \left.\frac{d E^2}{d x}\right|_0 = -\frac{j_{\e}}{\beta} \quad \text{and} \quad \left.\frac{d E^2}{d x}\right|_L = \frac{j_{\e}}{\beta}
\end{equation}
Figure \ref{fig:Current} (left) shows the electric field and the ion profiles for $L=1$ and $\beta^2=2$.  Due to the confined volume in which ions are generated, the associated current is only $j_{\e}=0.655$ as compared to two in the limit of large separations ($L\rightarrow \infty$).  Figure \ref{fig:Current} (right) shows the computed current density $j_{\e}$ as a function of the boundary separation $L$.  For $L\ll2$, the current is limited by the rate at which ions are generated within the electrolyte: $j_{\e} \approx \beta L/2$.  In dimensional units, the apparent conductivity of the electrolyte depends linearly the boundary separation as 
\begin{equation}
    \sigma = \frac{j_{\e}}{E_{\e}} \approx \frac{e k_{\r} n_{\eq}^2 L}{E_{\e}}  \quad \text{for} \quad L \ll 2\lambda
\end{equation}

\subsection{Semi-Infinite Electrolyte with a No-Flux Boundary}

\begin{figure}[b!]
    \centering
    \includegraphics[height=6cm]{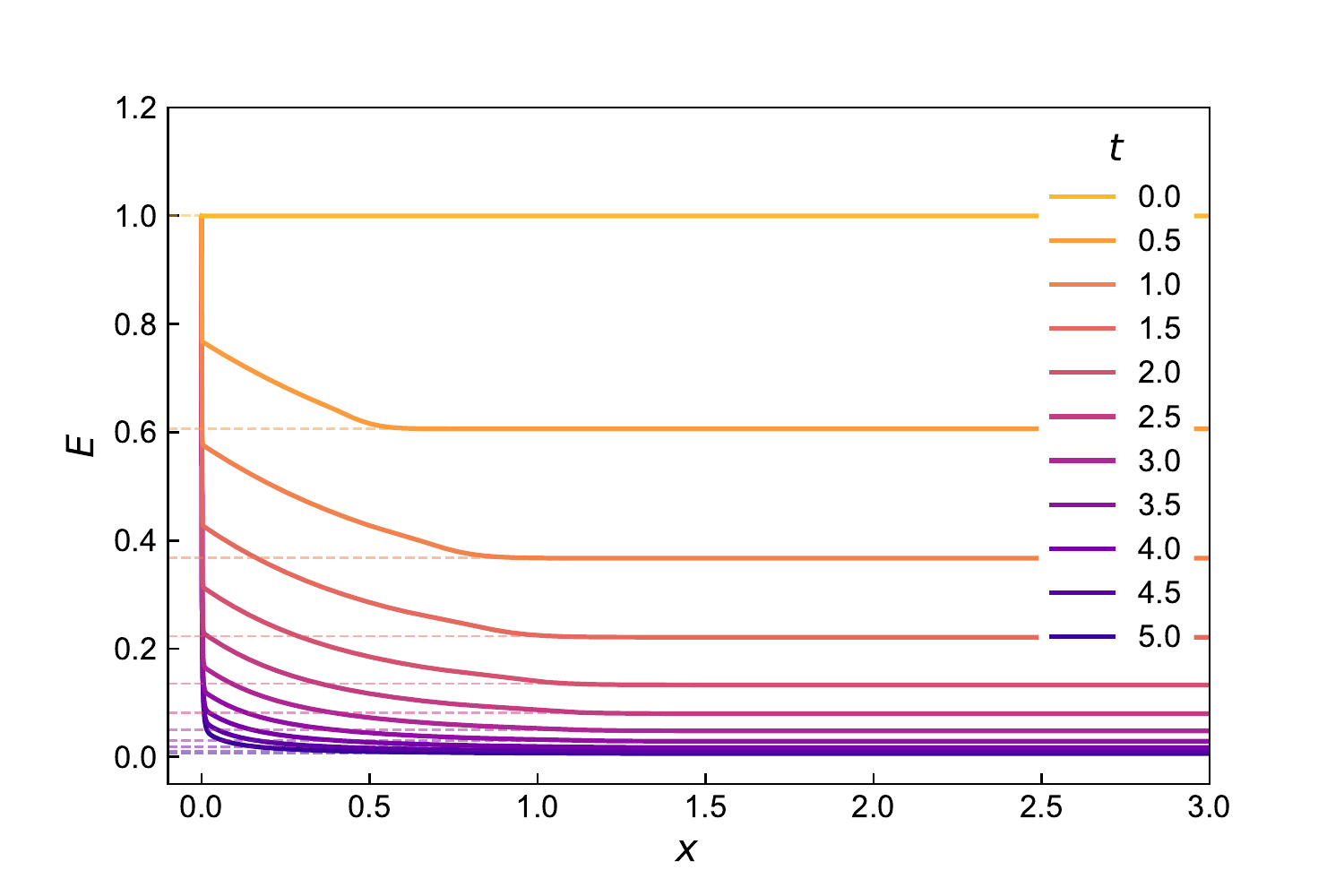}
    \caption{\textbf{One No-Flux Boundary.}  Transient electric field $E(x,t)$ near a no-flux boundary at $x=0$. The field was computed numerically for $\beta^2=2$ and $k_{\B}T/e\lambda E_{\e}=0.01$. All quantities are made dimensionless using the scales of equation (\ref{eq:E2_scales}): the field is scaled by the external field $E_{\e}$, position by the characteristic length $\lambda$, and time by the charge relaxation time $\varepsilon/2e^2\mu n_{\eq}$. The dashed line shows the predictions of equation (\ref{eq:relax}) that $E(x,t)=\exp(-t)$ for $x\gg 1$.}
    \label{fig:noFlux}
\end{figure}

We next consider a no-flux boundary at $x=0$ in contact with our model electrolyte on the domain $x>0$.  At the boundary, we specify the electric field $E_{\e}$, which drives the accumulation of charge within the diffusive boundary layer. For strong fields ($E_{\e} \gg k_{\B} T/e\lambda$), we can approximate this boundary layer by a region of vanishing thickness containing charge $q(t)$ per unit area 
\begin{equation}
    E(0,t) = E_{\e} + \frac{q(t)}{\varepsilon}
\end{equation}
The surface charge density evolves in time due to migration of ions as
\begin{equation}
    \frac{dq}{d t} = -e^2 \mu n_{\s}(0,t) E(0,t) = -\frac{e^2 \mu n_{\s}}{\varepsilon} (\varepsilon E_{\e} + q)
\end{equation}
For a field in the positive $x$-direction ($E_{e}>0$), we require that the concentration of positive ions vanish at the boundary as described by equation (\ref{eq:1D_bc1}). Far from the boundary, the ion concentrations approach their bulk values
\begin{equation}
    n_{\dd}(\infty,t) = 0 \quad \text{and} \quad n_{\s}(\infty,t) = 2 n_{\eq}
\end{equation}
Subject to these boundary conditions, equation (\ref{eq:1D}) describes the spatiotemporal evolution of the ion concentrations $n_{\pm}(x,t)$ and the electric field $E(x,t)$. 

At steady-state, the applied field $E_{\e}$ is fully screened by the charge accumulated at the boundary.  As a result, the ion concentrations and the field approach their initial values at long times
\begin{equation}
    q(\infty)=-\varepsilon E_{\e}, \quad n_{\dd}(x,\infty) = 0, \quad n_{\s}(x,\infty)=2n_{\eq}, \quad E(x,\infty) =0
\end{equation}
Small perturbations about this asymptotic solution decay as
\begin{equation}
    \frac{\partial n'_{\dd}}{\partial t} = - \frac{2 e^2 \mu n_{\eq}}{\varepsilon} n'_{\dd}, 
    \quad
    \frac{\partial n'_{\s}}{\partial t} = - 2 k_{\r} n_{\eq}  n'_{\s},
    \quad
    \frac{\partial q'}{\partial t} = - \frac{2 e^2 \mu n_{\eq}}{\varepsilon} q
\end{equation}
where $n'_{\dd}=n_{\dd}$, $n_{\s}'=n_{\s}-2n_{\eq}$, and $q'=q+\varepsilon E_{\e}$.  The rate $2k_{\r}n_{\eq}$ describes the characteristic rate of ion generation and recombination; the rate $2 e^2 \mu n_{\eq}/ \varepsilon$ is the charge relaxation rate of the electrolyte.  The ratio of these rates is the dimensionless parameter $\beta^2$ introduced above.

Figure \ref{fig:noFlux} shows the numerical solution for the case of diffusion-limited ion association with $\beta=2$. The field-induced boundary layer grows to a characteristic thickness $\lambda$ during a characteristic time $\lambda/e \mu E_{\e}$. Meanwhile, the bulk field outside of the boundary layer decays exponentially as
\begin{equation}
    E(x,t) = E_{\e} \exp\left(-\frac{2e^2 \mu n_{\eq} }{\varepsilon}t\right) \quad \text{for} \quad x\gg \lambda \label{eq:relax}
\end{equation}
Note that this result applies only when the electrolyte domain is thicker than the boundary layer thickness $\lambda$ for ion recombination.

\subsection{Finite Electrolyte between No-Flux and Absorbing Boundaries}

\begin{figure}[b!]
    \centering
    \includegraphics[height=6cm]{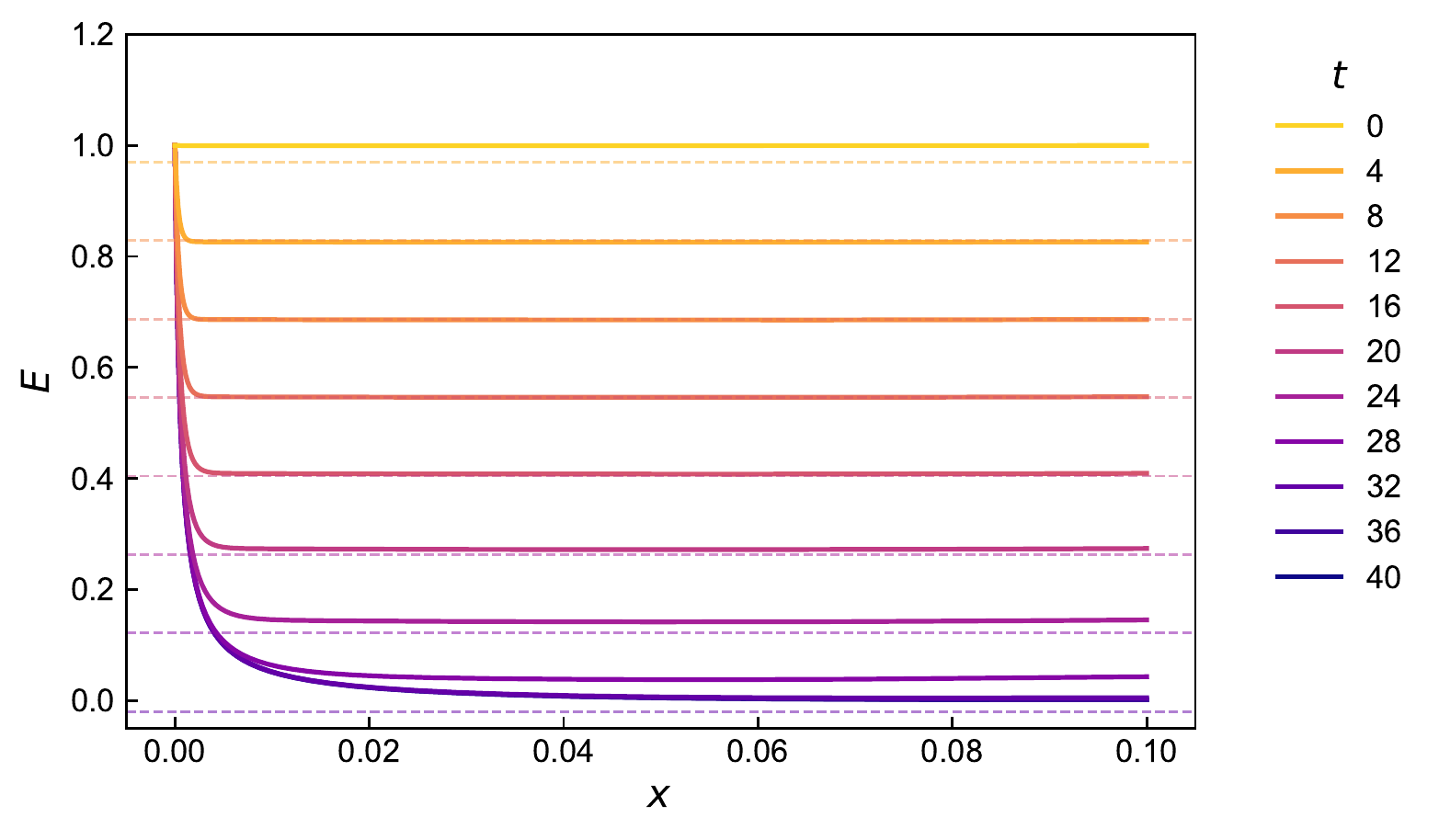}
    \caption{\textbf{Absorbing and No-Flux Boundaries.} Transient electric field $E(x,t)$ for an electrolyte confined between a no-flux boundary at $x=0$ and an absorbing boundary at $x=0.1$. The field was computed numerically for $\beta^2=2$ and $k_{\B}T/e\lambda E_{\e}=0.01$. All quantities are made dimensionless using the scales of equation (\ref{eq:E2_scales}): the field is scaled by the external field $E_{\e}$, position by the characteristic length $\lambda$, and time by the charge relaxation time $\varepsilon/2e^2\mu n_{\eq}$. The dashed lines show that the field decreases at a constant rate $\beta L/4$ as predicted by equation (\ref{eq:charge}).}
    \label{fig:Absorb_NoFlux}
\end{figure}

Finally, we consider our model electrolyte confined between an no-flux boundary at $x=0$ and an absorbing boundary at $x=L$ (Figure \ref{fig:Absorb_NoFlux}). This configuration is similar to the gap separating a solid particle (no-flux) from the electrode surface (absorbing).  When the gap is small relative to the recombination length ($L\ll \lambda$), the rate of charge accumulation on the no-flux boundary is limited by the finite rate of ion generation within the electrolyte. Following a short time during which the initial charge carriers are removed ($\Delta t \sim L/e\mu E_{\e}$), the charge on the no-flux boundary increases linearly with time as 
\begin{equation}
    q(t) = e k_{\r} n_{\eq}^2 L  t \label{eq:charge}
\end{equation}
resulting in the gradual attenuation of the applied field. This charging process continues for a duration $\Delta t \sim \varepsilon E_{\e} / e k_{\r} n_{\eq}^2 L$, at which point the applied field is fulled screened.  Importantly, the charging time for the confined electrolyte is much slower than that of the unbounded electrolyte, which is equal to the charge relaxation time $\varepsilon / 2e^2\mu n_{\eq}$.  The ratio between these times is $4\lambda / \beta L \gg 1$  (or, equivalently, $2\ell / L \gg 1$).

\clearpage
\section{Leaky Dielectric Model}

We consider a solid dielectric sphere of radius $a$ immersed in a model electrolyte (Section \ref{sec:model_electrolyte}) at a distance $\delta$ from a plane electrode. The application of an external field $\ve{E}_{\e}$ normal to the electrode drives the migration of charge carriers, which accumulate at the particle-fluid interface. For strong fields $E_{\e} e \lambda/ k_{B} T\gg 1$ and large particles $a/\lambda\gg 1$, the majority of this charge is confined within a thin region of thickness $k_{\B} T/e E_{\e}$ at the particle surface (see Fig.\ \ref{fig:noFlux}). This charged region is assumed to lie within the idealized surface of shear that separates the rigid particle from its fluid surroundings; consequently, the charge moves in lock-step with the particle as it rotates. Following Melcher \& Taylor,\autocite{Melcher1969,Saville1997} we neglect any free charge remaining outside of the shear plane---for example, that in the boundary layer of thickness $\lambda$.  In doing so, we neglect also any electrokinetic flows that might originate from the action of the electric field on this charge. 

With these simplifications, the electric field $\ve{E}_{\f}$, fluid velocity $\ve{u}$, and pressure $p$ within the fluid are governed by 
\begin{align}
    \nabla\cdot(\varepsilon_{\f} \ve{E}_{\f}) &= 0 \label{eq:leaky_Ef}
    \\
    -\nabla p + \eta \nabla^2 \ve{u} &=0
    \\
    \nabla \cdot \ve{u} &= 0
\end{align}
Note that we have omitted inertial contributions to the Navier-Stokes equations owing to the small Reynolds numbers characterizing particle motion. The dielectric particle moves as a rigid body with angular velocity $\ve{\Omega}$; the electric field within the particle is governed by 
\begin{equation}
    \nabla\cdot(\varepsilon_{\p} \ve{E}_{\p}) = 0 \label{eq:leaky_Ep}
\end{equation}

At the interface between the particle and the fluid (denoted $r=a$), the electric potential is continuous while the normal component of electric displacement changes discontinuously by an amount equal to the surface charge density $q$  
\begin{equation}
    \begin{drcases}
    \phi_{\p} = \phi_{\f} \label{eq:leaky_bc_phi}
    \\
    \ve{n}\cdot (\varepsilon_{\p} \ve{E}_{\p} - \varepsilon_{\f} \ve{E}_{\f}) = q 
    \end{drcases}  \quad \text{for } r = a
\end{equation}
where $\ve{n}$ is the unit normal vector directed into the fluid. The surface charge evolves in time due both to the rotation of the sphere and to the flow of charge from the surrounding fluid 
\begin{equation}
    \frac{\partial q}{\partial t} + \nabla_{\s}\cdot (q \ve{u}) + \ve{n}\cdot \ve{j}_{\e} = 0 \quad \text{for } r = a
\end{equation}
where $\ve{j}_{\e} = \sigma_{\f} \ve{E}_{\f}$ is the electric current in the fluid, $\ve{u}=\ve{\Omega}\times a \ve{n}$ is the velocity of the sphere surface, and $\nabla_{\s}$ is the surface divergence. Here, the conductivity of the dielectric particle is assumed to be zero. At the electrode surface (denoted $\mathcal{S}_{\e}$), the electric potential $\ve{\phi}$ is zero as is the fluid velocity
\begin{equation}
    \begin{drcases}
    \phi_{\f} = 0
    \\
    \ve{u} = 0
    \end{drcases}  \quad \text{for } \ve{r}\in \mathcal{S}_{\e}
\end{equation}
Far from the sphere, the field approaches a constant value and the fluid velocity approaches zero
\begin{equation}
    \begin{drcases}
    \ve{E}_{\f} = \ve{E}_{\e}
    \\
    \ve{u} = 0
    \end{drcases}  \quad \text{for } r\rightarrow \infty \label{eq:leaky_bc_infty}
\end{equation}

Rather than solve the Stokes equation directly for the fluid velocity, we use the resistance matrix formulation\autocite{Kim2005} to relate the angular velocity of the sphere to the electric torque as 
\begin{equation}
    \ve{R} \cdot \ve{\Omega} = \ve{L}_{\e}
\end{equation}
where $\ve{R}$ is the resistance matrix for a sphere rotating above a solid plane.\autocite{Goldman1967} The relevant components have the form $R_{xx}=R_{yy}=8\pi\eta a^3 f(\delta/a)$ where $f(\delta/a)$ is a dimensionless coefficient that depends on the surface separation $\delta$ scaled by the particle radius. The electric torque on the particle is given by
\begin{equation}
    \ve{L}_{\e}= \oint \ve{r} \times (\ve{n}\cdot \ve{\Sigma}_{\e})~dS
\end{equation}
where the integral is taken over a spherical surface enclosing the particle and its surface charge $q$. Here, the electric stress tensor within the fluid is defined as $\ve{\Sigma}_{\e}=\varepsilon_{\f} \ve E_{\f} \ve E_{\f} - \tfrac{1}{2}\varepsilon_{\f} (\ve E_{\f} \cdot \ve E_{\f}) \ve I$.

In previous works,\autocite{Das2013, Hu2018} the governing equations (\ref{eq:leaky_Ef})--(\ref{eq:leaky_Ep}) together with boundary conditions (\ref{eq:leaky_bc_phi})--(\ref{eq:leaky_bc_infty}) have been solved using multipole expansion approaches, which generally truncate at the dipole moment level. In our case, the particle is close to a planar electrode, and the effects of higher order moments must be taken into account. We therefore solve the governing equations self-consistently with the finite element method.

All numeric simulations are performed using the commercial finite-element software COMSOL. The Poisson equation is solved by the electrostatic module of COMSOL; the surface charge conservation equation is implemented as a user-defined partial differential equation via the boundary PDE module of COMSOL. Since the surface charge conservation equation is a partial differential equation defined on a 2D surface, the surface divergence operator $\nabla_s \cdot$ needs to be the covariant divergence, which depends on the metrics of the surface. For the convenience of numeric implementation, the surface charge conservation equation of a spherical surface can be expressed in the Cartesian coordinate as:
\begin{equation}
    \frac{\partial q}{\partial t}+\ve n \cdot (\sigma_{\f}\ve E_{\f}) + \ve u \cdot \nabla q=0
\end{equation}

\clearpage
\section{Oscillations of Cylindrical Particles}
\vspace{1cm}

\begin{figure}[h]
   \centering
    \includegraphics{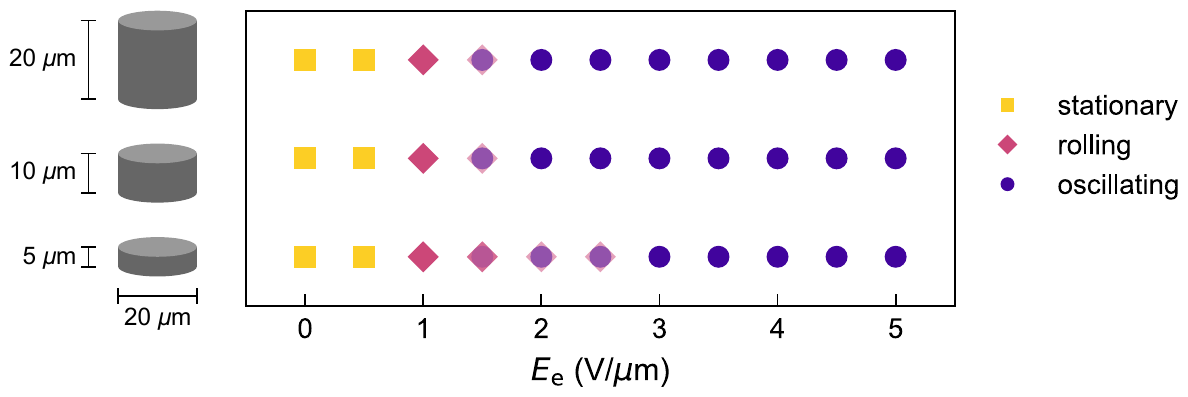}
    \caption{Phase diagram showing the observed dynamics for cylindrical particles of different aspect ratios as a function of the external field strength $E_{\e}$. Plotted markers denote the three observed behaviors: stationary, rolling, and oscillating. The illustrations on the left show the dimensions of the cylindrical particles, which were fabricated from SU-8 photoresist by photolithography. Cylinders with 1:1 aspect ratio rotate about their axis of symmetry; the other cylinders (1:2 and 1:4 aspect ratios) rotate about an axis normal to their symmetry axis.}
    \label{fig:othershape}
\end{figure}

\clearpage
\printbibliography